\documentclass[a4paper,10pt]{article}
\usepackage[pdftex]{graphicx}
\PassOptionsToPackage{hyphens}{url}\usepackage{hyperref}
\usepackage{amsmath,amsfonts,amssymb}
\usepackage{fullpage}
\usepackage{authblk}
\usepackage{setspace}
\usepackage{subcaption}
\usepackage{booktabs}
\usepackage{tabularx}

\usepackage[explicit]{titlesec}
\usepackage{sidecap}
\usepackage{pbox}
\usepackage[superscript]{cite}

\usepackage{amsfonts}
\usepackage{amsmath}
\usepackage{multirow}
\usepackage{longtable}
\usepackage{graphicx}
\usepackage{xcolor}
\newcommand{\edit}[1]{{#1}}

\date{}
\title{\bf{Dark Web Marketplaces and COVID-19: before the vaccine}}

\author[1]{Alberto Bracci}
\author[1]{Matthieu Nadini}
\author[2]{Maxwell Aliapoulios}
\author[2]{Damon McCoy}
\author[3]{Ian Gray}
\author[4,5]{Alexander Teytelboym}
\author[6]{Angela Gallo}
\author[1,7,8,*]{Andrea Baronchelli}

\affil[1]{\small City, University of London, Department of Mathematics, London EC1V 0HB, UK}
\affil[2]{\small Center for Cybersecurity (CCS), New York Univ. Tandon School of Engineering, Brooklyn, NY 11201, USA}
\affil[3]{\small Global Intelligence Team, Flashpoint. New York, NY 10003, USA}
\affil[4]{\small Institute for New Economic Thinking, Oxford Martin School, University of Oxford, Oxford OX2 6ED, UK}
\affil[5]{\small Department of Economics, University of Oxford, Oxford OX1 3UQ, UK}
\affil[6]{\small Department of Finance, Cass Business School, London EC1Y 8TZ, UK }
\affil[7]{\small UCL Centre for Blockchain Technologies, University College London, UK}
\affil[8]{\small The Alan Turing Institute, British Library, 96 Euston Road, London NW12DB, UK}

\affil[*]{\small Corresponding author: Andrea.Baronchelli.1@city.ac.uk}

\begin{document}
\maketitle

\textbf{The COVID-19 pandemic has reshaped the demand for goods and services worldwide. The combination of a public health emergency, economic distress, and misinformation-driven panic have pushed customers and vendors towards the shadow economy. In particular, Dark Web Marketplaces (DWMs), commercial websites \edit{accessible} via free software, have gained significant popularity. Here, we analyse \edit{851,199} listings extracted from \edit{30} DWMs between January 1, 2020  and \edit{November 16}, 2020. We identify \edit{788} listings directly related to COVID-19 products and monitor the temporal evolution of product categories including \textit{Personal Protective Equipment} (PPE), \textit{medicines} (e.g., Hydroxyclorochine), and \textit{medical frauds}. Finally, we compare trends in their temporal evolution with variations in public attention, as measured by Twitter posts and Wikipedia page visits. We reveal how the online shadow economy has evolved during the COVID-19 pandemic and highlight the importance of a continuous monitoring of DWMs, especially \edit{now that} real vaccines \edit{are} available and in short supply. We anticipate our analysis will be of interest both to researchers and public agencies focused on the protection of public health.}

\vspace{1cm}

\section*{Introduction}

COVID-19 gained global attention when China suddenly quarantined the city of Wuhan on January 23, 2020~\cite{nytimescoronavirus}. Declared a pandemic by the World Health Organization on March 11, 2020, at the moment of writing the virus has infected more than \edit{62,000,000} people and caused over \edit{1,450,000} deaths worldwide~\cite{world2020coronavirus}. Measures such as social distancing, quarantine, travel restrictions, testing, and tracking have proven vital to containing the COVID-19 pandemic~\cite{kraemer2020effect}. 

Restrictions have shaken the global economy and reshaped the demand for goods and services worldwide, with an estimated $2.5-3\%$ \edit{world GDP} loss since the crisis started~\cite{fernandes2020economic}. Demand for many products has fallen; for example, \edit{the price of Brent crude oil} decreased from 68.90 USD a barrel on January 1, 2020 to 43.52 USD as of August 2, 2020~\cite{arezki20204,BloombergOilPrice}. Meanwhile demand for other products, \edit{such as} toilet paper ~\cite{toilet_paper}, dramatically increased. As a result of increased demand, some products \edit{have been} in short supply. Individual protective masks were sold in the United States at $7$ USD on February 2, 2020~\cite{markshortage} and the price of alcohol disinfectant doubled on July 1, 2020 in Japan~\cite{disinfectantdoubleprice}. \edit{Additionally}, anti-gouging regulations were introduced to control prices, which significantly affected the public attention on products related to COVID-19~\cite{chakrabortianti}.
%, like hand sanitisers and toilet papers~\cite{chakrabortianti}. 
As this trend has continued, further exacerbated by online misinformation, numerous customers have sought to fulfill their needs through illicit online channels~\cite{GIATOC, EuropolEMCDDA}. 

\edit{Dark Web Marketplaces (DWMs)} offer \edit{access} to the shadow economy \edit{via specialized browsers}, like Tor~\cite{dingledine2004tor}. %These online
DWMs offer a variety of goods including drugs, firearms, credit cards, and fake IDs~\cite{GwernDarkNets}. The most popular currency on DWMs is Bitcoin~\cite{nakamoto2008bitcoin}, but other cryptocurrencies are accepted for payment as well. The first modern DWM was the Silk Road, launched in 2011~\cite{christin2013traveling} and shut down by the FBI~\cite{SILKROADSEALED} in 2013. Since then, dozens more DWMs have sprung up and many have shut down due to police action, hacks, or scams. Today, DWMs form an ecosystem~\cite{soska2015measuring} that has proven extremely resilient to law-enforcement. Whenever a DWM is shut down, users swiftly migrate to alternative active DWM and the economic activity recovers within a matter of days~\cite{elbahrawy2019collective}.

\edit{Researchers have studied DWMs since the emergence of Silk Road}~\cite{christin2013traveling}\edit{, through a series of case studies~\cite{van2013silk, van2014responsible, lacson201621st}, and comparative analyses~\cite{barratt2014use, martin2014lost, aldridge2014not, dolliver2015criminogenic, dolliver2015evaluating, broseus2016studying}. Past efforts have mostly focused on specific goods, such as drugs or digital products~\cite{rhumorbarbe2016buying}. However, these studies experienced technical difficulties in data collection preventing researchers from analysing a large and up-to-date dataset. As a result, several questions remain open, among which are:
\begin{itemize}
\item how do DWMs react to sudden shocks (e.g., shortages) in the traditional economy? 
\item how do DWMs respond to trends in public attention?
\end{itemize}
}

In this study, we \edit{address these questions by analysing a new, large, and up-to-date dataset. We studied} \edit{851,199} listings extracted from \edit{30} DWMs between January 1, 2020 and \edit{November 16}, 2020, \edit{right before the first worldwide vaccination campaign started in the United Kingdom~\cite{vaccination_uk}}. We identify \edit{788} COVID-19 specific listings that range from protective masks~\cite{MasksCOVID} to hydroxychloroquine medicine~\cite{covid19_medicines}. These listings were observed \edit{9,464} times during this period, allowing us to investigate their temporal evolution. We compare this COVID-19 related shadow economy with public attention measured through Twitter posts (tweets)~\cite{chen2020tracking} and Wikipedia page visits~\cite{WikimediaAPI}. Finally, we inspect listings that mentioned delays in shipping or sales because of COVID-19. \edit{We} significantly extend previous analyses that surveyed $222$ COVID-19 specific listings extracted from $20$ DWMs on a single day (April $3$, $2020$)~\cite{broadhurstavailability} and, to the best of our knowledge, offer the most comprehensive overview of the DWM activity generated by the ongoing pandemic. 

\edit{We found that DWMs promptly respond to signals coming from the traditional economy, increasing or decreasing the offer of goods according to their availability on \edit{regulated} markets. For example, protective masks appeared in DWMs in March, when they were in short supply in the \edit{regulated} economy, and became more \edit{scarce on DWMs} later on when masks could be easily bought in shops. We also found that DWMs swiftly react to changes in public attention as measured through Twitter posts and Wikipedia page views. Finally, we registered spikes in the number of listings mentioning COVID-19 in correspondence with lockdown measures in March and October. Our results are of interest to different audiences: the academic community may further explore the behaviour of DWMs in relation to social changes. Policy makers can better understand the effects that new legislation have in the shadow economy. Practitioners may learn that DWMs posit additional threats to public health. A finding especially important nowadays due to the production of COVID-19 vaccines.}

\edit{The manuscript is organized as follows. In the Background Section, we introduce DWMs with a brief overview of their history. In the Data Section, we explain how we obtained our DWMs, Twitter, and Wikipedia datasets. The main outcomes of our work are presented in the Results Section, while in the Discussion Section we compare them with the established technical literature. Finally, in the Conclusion Section, we highlight the contributions of our work that are relevant to different audiences as well as future research developments.}

\section*{Background: Dark Web Marketplaces}
\label{sec:background}

The online shadow economy is as old as the Internet. The first reported illegal online deal involved drugs and took place in 1972~\cite{markoff2005dormouse}. The World Wide Web~\cite{berners1994world} facilitated the emergence of online illicit marketplaces~\cite{HiveWWW,FarmerMarketWWWTor} but the first marketplaces could not guarantee anonymity and \edit{therefore permitted} the traceability of users by law enforcement~\cite{barratt2011discussing}.

Modern DWMs originated and still operate online, but outside the World Wide Web in an encrypted part of the Internet whose contents are often not indexed by standard web search-engines~\cite{martin2014drugs}. \edit{\textit{Silk Road} marketplace, which launched in 2011, was the first modern DWM~\cite{christin2013traveling}. It proposed a new way of trading drugs and other illegal products online and anonymously~\cite{barratt2014use, martin2014lost, aldridge2014not}. There were two key ingredients of Silk Road's success. \edit{First,} potential customers could access it using the Tor browser~\cite{dingledine2004tor}, which made their traceability difficult. \edit{Second,} purchases were made in Bitcoin~\cite{nakamoto2008bitcoin}, which provided a degree of privacy to buyers and sellers~\cite{cappa2020collecting, masoni2016darknet, rhumorbarbe2016buying}.} After the FBI shut down Silk Road in 2013~\cite{SILKROADSEALED}, new DWMs quickly appeared, offering drugs, firearms, credit cards, and fake IDs~\cite{GwernDarkNets}. These DWMs also adapted to further increase the level of privacy and security offered to users~\cite{van2013silk, van2014responsible}, such as the Invisible Internet Project (I2P)~\cite{zantout2011i2p} and escrow checkout services~\cite{wehinger2011dark}. \edit{Tor, now available for mobile devices as well, still offers more privacy than many other popular mobile applications~\cite{hayes2020effective} \edit{and Bitcoin is currently the most popular cryptocurrency in DWMs \cite{lee2019cybercriminal, foley2019sex, moser2018empirical}.}} 

Trade today \edit{on DWMs} is worth at least several hundreds of millions of USD per year, and involves hundreds of thousands of buyers and vendors~\cite{SILKROADSEALED, soska2015measuring, elbahrawy2019collective,  Christin2017AnEA, WallStreetMarket, FBIAlphabay}. As a result, law enforcement agencies have put considerable effort into combating them~\cite{SILKROADSEALED,WallStreetMarket, FBIAlphabay}. Furthermore, DWMs have been targets of cybercriminal actors through use of distributed denial-of-service (DDoS) attacks, hacking attempts\edit{, and some DWMs} also shut down due to administrators stealing funds from customers directly~\cite{exit_scam_1,exit_scam_2}. However, DWMs have organised into a robust ecosystem which has proven exceptionally resilient to closures and whenever a DWM is closed, the users trading higher volumes of Bitcoins migrate to active DWMs or establish new ones~\cite{elbahrawy2019collective}. 

The resilience and functioning operations of modern DWMs are possible partially because of numerous websites and forums where users can share their experience\edit{s}. One example is Dread~\cite{Dread}, a Reddit-like forum created in 2018 after the closure of the dedicated pages on Reddit~\cite{subreddits_closure}. Other ad-hoc platforms exist to monitor whether known DWMs are active or currently unavailable~\cite{OnionLive, DarkNetLive, DarkFail, DNStats}. \edit{Additional mechanisms, like feedback and ratings systems~\cite{soska2015measuring}, enhance the resilience of these DWMs and build trust towards the DWM and its vendors}. 
% Mechanisms to enhance the resilience of these DWMs and build trust towards the DWM and its vendors include feedbacks and ratings~\cite{soska2015measuring}. 
\edit{In a similar way to \edit{regulated} online marketplaces, DWM buyers \edit{are asked} to leave feedback and a rating after a purchase.}
% On most DWMs, buyers have to leave feedback and a rating after a purchase, similarly to what happens on \edit{regulated} online marketplaces.
Additionally, DWM administrators often act as vendor moderators by banning vendors or specific categories of products. \edit{Some examples} \edit{of this} \edit{are} DarkBay, where banned categories include human trafficking, contract killing and weapons~\cite{DarkNetLive_ban}, \edit{and} Monopoly marketplace, where COVID-19 fake vaccine listings were recently banned by moderators~\cite{monopoly_ban}.

\edit{DWMs have been used to circumvent laws and regulations. They have been the subject of many case studies~\cite{van2013silk, van2014responsible, lacson201621st} and comparative analyses~\cite{barratt2014use, martin2014lost, aldridge2014not, dolliver2015criminogenic, dolliver2015evaluating, broseus2016studying}. These studies highlighted that illicit online transactions in DWMs are perceived as safer than negotiating in face-to-face drug markets~\cite{van2013silk}. They are based on the concept of ``harm reduction,'' where vendors prefer to sell tested and high quality products~\cite{van2014responsible}. \edit{Although} DWMs form an online community, \edit{they are} made unstable by \edit{their} profit-based mentality of capitalism~\cite{lacson201621st}. Vendors customize their products to match the specialisation of different DWMs thus creating an efficient distribution network~\cite{broseus2016studying}, which sometimes goes beyond a base retail market~\cite{dolliver2015evaluating}. While these characteristics favour the DWM economy against the offline shadow economy, DWMs sell a variety of illicit products~\cite{barratt2014use, martin2014lost, aldridge2014not, dolliver2015criminogenic}, such as, drugs, fake IDs, ``how to'' manuals (for scams, bombs etc.), and weapons. One prominent category is that of digital goods~\cite{digital_goods_leuven}, including \edit{ransomware}, social engineering guides, \edit{and} financial \edit{malware} to steal credit cards and bank account credentials.}

It is hard to estimate how many live DWMs currently exist. \edit{Some} recent reports include \edit{one from} independent researcher Gwern, \edit{who} identified 19 live platforms on April 22, 2020~\cite{gwern_live_markets}, the website darknetstats, which registered 10 live ``established'' DWMs on May 27, 2020~\cite{darknetstats_live_markets}, and \edit{one where 20 DWMs were observed one single day (April 3, 2020)~\cite{broadhurstavailability}}. Currently established DWMs \edit{at the time of writing} include \edit{Hydra and White House} marketplaces.

\section*{Data and methods}

\begin{figure}[h!]
  \centering
  \includegraphics[width=12cm]{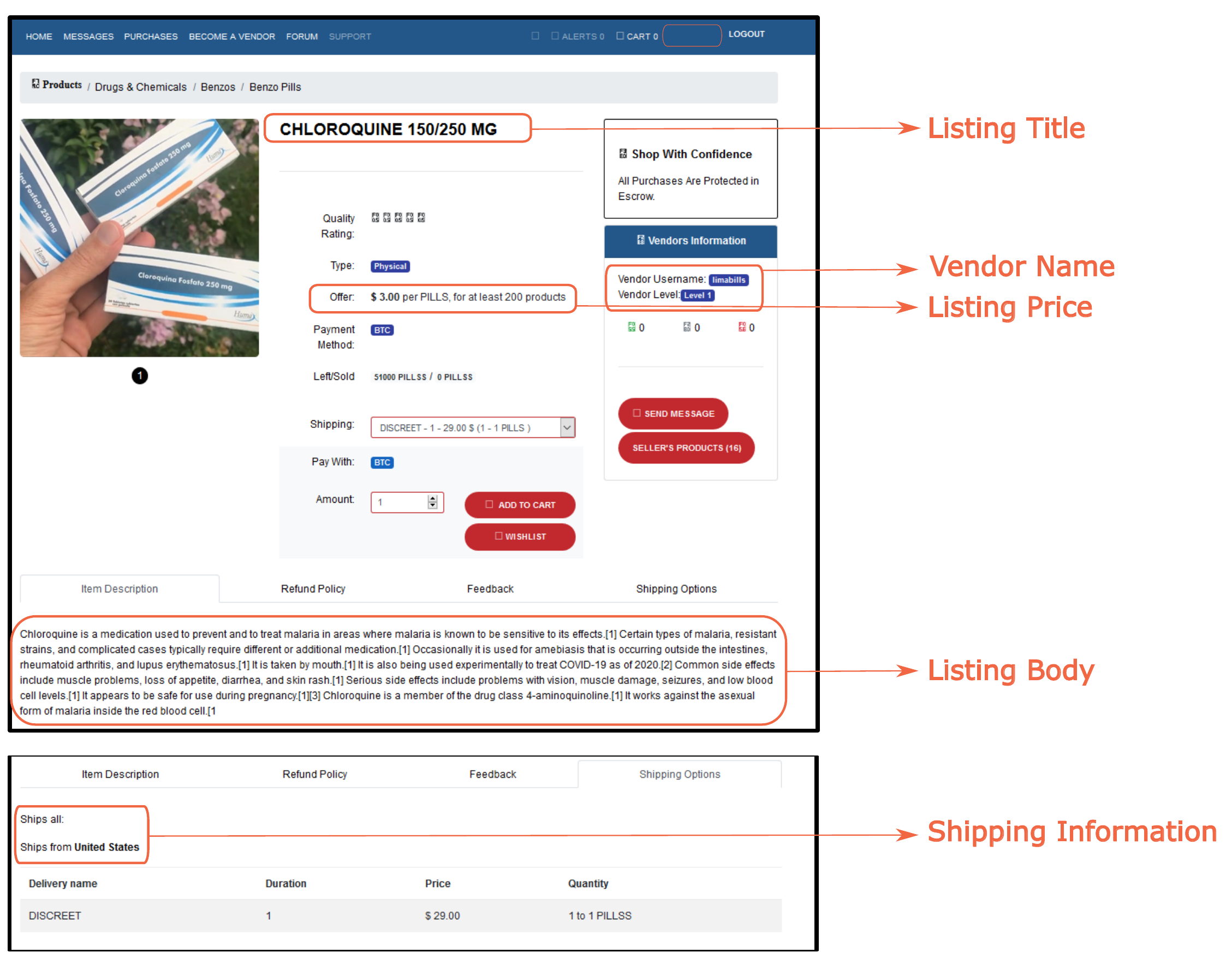}
  \caption{Example of a DWM listing. Screenshot of a chloroquine listing in the DarkBay/Dbay marketplace, where we highlight some of its salient attributes. Among the attributes considered in this work and shown in Table~\ref{Listing_attributes}, ``Time'' and ``Marketplace name'' attributes are not present in this screenshot, while the ``Quantity'' attribute is not fixed by the vendor.}
  \label{From_listings_to_data}
\end{figure}

\subsection*{Dark web marketplaces}

The listings used for our study were obtained by web crawling DWMs. Web crawling consists of extracting data from websites and is performed by specialized software. Web crawling DWMs is a challenging task because crawlers must bypass several protective layers. Most DWMs require authentication and some even require a direct invitation from a current member. \edit{Additionally,} strong CAPTCHAs~\cite{ball2019data} are implemented to avoid otherwise easy and automated access to the online DWM. Several research groups tried to overcome these challenges. Some \edit{examples are}, HTTrack software used in~\cite{christin2013traveling}, a combination of \textit{PHP}, the \textit{curl} library, and \textit{MySQL} was proposed in~\cite{baravalle2016mining}, the Python library \textit{scrapy} adopted in~\cite{celestini2017tor}, and an automated methodology using the \textit{AppleScript} language utilized in~\cite{hayes2018framework}. There are currently very few open source tools available~\cite{decary2013datacrypto, ball2019data} \edit{for crawling DWMs}, \edit{which} often \edit{leaves} companies and federal agencies to rely on commercial software~\cite{xbyte}. Downloading content from DWMs remains a challenging task, \edit{and the objective} becomes even harder when the research study requires monitoring multiple DWMs for an extended period of time. 

Our dataset contains listings crawled \edit{from} \edit{$30$} DWMs between January 1, 2020 and \edit{November 16, 2020} by Flashpoint Intelligence~\cite{flashpoint}\edit{, a company specializing in online risk intelligence}. \edit{It includes the most popular DWMs in 2020, such as Hydra, White House, Empire and Dark Market~\cite{darknetstats_live_markets, broadhurstavailability}}. \edit{The crawlering pipeline} consists of \edit{authenticating into} DWMs and downloading key attributes for each active listing, as highlighted in Figure~\ref{From_listings_to_data}. Each DWM was crawled for at least $90$ different days. We categorized the COVID-19 specific listings into \textit{PPE}, \textit{medicines}, \edit{\textit{guides on scamming}, \textit{web domains}, } \textit{medical frauds}, \textit{tests}, \edit{\textit{fake medical records}}, and \textit{ventilators}. \edit{Representative examples of listings relative to these categories are presented in Table~\ref{Categories_Keywords}, with specific listing examples in Appendix~\ref{Examples_of_listings_specific}}. Only a fraction of the selected listings were actual COVID-19 specific listings, since mitigation measures to prevent COVID-19 spreading have also impacted illegal trades of other listings. For instance, a vendor might sell cocaine and mention shipping delays due to COVID-19. We included such cases in the category COVID-19 \textit{mentions}. For details about data pre-processing, see Appendix~\ref{Data_preprocessing}, where we explain how we select listings related to COVID-19 and how we classify them in categories. We remark that our pre-processing pipeline is biased towards the English language, and this constitutes a limitation of our work.

\begin{table}[ht]
  \centering
\caption{Categories used to classify the selected COVID-19 dataset. The first five categories constitute COVID-19 specific listings, while the last one, called COVID-19 \textit{mentions}, includes listings mentioning one of the keywords in Table~\ref{Keywords} without selling actual COVID-19 specific listings.}
      \begin{tabular}{ll}
        \hline
        Category & Examples \\
        \hline
        PPE & gloves, gowns, masks, n95 \\
        Medicines & azithromycin, chloroquine, azithromycin, favipiravir, remdesivir\\
        \edit{Guides on scamming} & \edit{how to illicitly get COVID-19 relief packages} \\
        \edit{Web Domains} & \edit{covid-testing.in, coronavintheworld.com}\\
        Medical Frauds & antidotes, vaccines, allegedly curative recreational drug mixes \\
        Tests & diagnosis, test \\
        \edit{Fake Medical Records} & \edit{medical record, medical certification} \\
        Ventilators & medical ventilators \\
        COVID-19 mentions & computer, drugs, scam (excluding listings in the \edit{previous} categories) \\
        \hline
      \end{tabular}
\label{Categories_Keywords}
\end{table}

Overall, our dataset includes a total of \edit{851,199} unique listings, which \edit{were} observed a total of \edit{8,538,593} times \edit{between January 1, 2020 and November 16, 2020}. In Table~\ref{BasicStatistics} we report the breakdown of the number of unique listings and their total observations in each of the \edit{30} DWMs. \edit{We did not find any mention of COVID-19 on} \edit{12} DWMs (\edit{Atshop,} Black Market Guns, Cannabay, Darkseid, ElHerbolario, \edit{Exchange,} Genesis, \edit{Mouse in Box,} Rocketr, Selly, Skimmer Device and Venus Anonymous). This makes sense as these DWMs are primarily focused on specific goods with \edit{a pre-defined} listing text structure. A brief description of each DWM together with its specialization can be found in Table~\ref{market_details}. On the remaining 18 DWMs, there were a total of \edit{10,455} unique listings related to COVID-19, which constitutes less than $1\%$ of the entire dataset. These listings were mostly composed of drugs that reported discounts or delays in shipping due to COVID-19. Listings concerning more specific COVID-19 goods such as \textit{masks}, \textit{ventilators}, and \textit{tests} were available on \edit{13} DWMs (Connect, \edit{Cypher, }DarkBay/DBay, DarkMarket, Empire, \edit{Hydra, MagBO, Monopoly, Plati.market, Torrez, }CanadaHQ, White House, and Yellow Brick). There were \edit{788} total COVID-19 specific listings in these DWMs which were observed \edit{9,464} times during the analysed time period. 

\begin{table}[ht]
  \centering
\caption{This table reports the number of days each marketplace was crawled, the number of unique listings, all and COVID-19 specific, and the number of listing observations, all and COVID-19 specific. CanadianHQ indicates The Canadian HeadQuarters marketplace. }
      \begin{tabular}{lccccc}
        \hline
          \multirow{2}{*}{Name marketplace} & Days & Listings  & Listings & Observations & Observations  \\ 
         & crawled & All	& COVID-19 specific & All & COVID-19 specific \\ \hline
Black Market Guns  & 	163  & 	18 & 0 & 	2,934 &  0\\
CanadaHQ & 	94 & 	21,853 & 3 & 	145,202 & 53\\
Cannabay & 119 & 	1,074 & 0 & 	1,303 & 0\\
Cannazon & 	100 & 	2,760 & 0 & 	4,606 & 0\\
Connect & 179 & 	476 & 2 & 	13,579 & 23\\
DarkBay/DBay & 	127 & 	105,921 & 421 & 	554,535 & 6570\\
DarkMarket & 	92 & 	32,272 & 19 & 	37,742& 20\\
Darkseid & 	189 & 	8 & 0 & 	1,512 & 0\\
ElHerbolario & 	186 & 	13 & 0 & 	1,430 & 0\\
Empire & 	107 & 	26,010 & 33 & 	93,163 & 46\\
Genesis & 	188 & 	216,792 & 0 & 	2,174,217 & 0\\
Hydra & 	189	 & 297 & 0 & 	37,665 & 0\\
MEGA Darknet & 	135 & 	754 & 0 & 	1,596 & 0\\
Plati.Market & 	189 & 	11,678 & 0 & 	17,214 & 0\\
Rocketr & 	189 & 	460 & 0 & 	7,843 & 0\\
Selly & 	91 & 	462 & 0 & 	1,523 & 0\\
Shoppy.gg & 	189 & 	8,412 & 0 & 486,819 & 0\\
Skimmer Device & 	189 & 	12 & 0 & 	2,268 & 0\\
Tor Market & 	130 & 	634 & 0 & 	25,328 & 0\\
Venus Anonymous & 	177 & 	84 & 0 & 	14,644 & 0\\
White House & 96 & 	21,377 & 5 & 	320,360 & 118\\
Willhaben & 	189 & 	14,626 & 0 & 	45,774 & 0\\
Yellow Brick & 	117 & 	6,379 &  33 & 	97,583 & 329\\ \hline
       Total & $>$ 90 & 472,372	& 518 &  4,088,840 & 7,159 \\ \hline
      \end{tabular}
\label{BasicStatistics}
\end{table}

\subsection*{Twitter}

We sampled tweets related to COVID-19 using a freely available dataset introduced in Chen et al~\cite{chen2020tracking}. We downloaded the tweets ID from the public GitHub repository \edit{and then} used the provided script to recover the original tweets through the Python library \textit{twarc}. We studied the temporal evolution of the number of tweets mentioning selected keywords, like ``chloroquine''. In line with our dataset of DWM listings, most of the tweets considered were written in English and the time period considered ranges from January 21, 2020 to \edit{November 13}, 2020. 

\subsection*{Wikipedia}

We used the publicly available Wikipedia API~\cite{WikimediaAPI} to collect data about the number of visits at specific pages related with COVID-19, like chloroquine. The Wikipedia search engine was case-sensitive and we considered strings with the first letter uppercase, while the others lowercase. We looked for the number Wikipedia page visits in the English language from January 1, 2020 to \edit{November 16}, 2020.

\section*{Results}

We assessed the impact of COVID-19 on online illicit trade along three main criteria. First, we focused on the \edit{13} DWMs containing at least one COVID-19 specific listing, analysing their offers in terms of the categories \textit{PPE}, \textit{medicines}, \edit{\textit{guides on scamming}, \textit{web domains}, } \textit{medical frauds}, \textit{tests}, \edit{\textit{fake medical records}}, and \textit{ventilators}, as introduced in Table~\ref{Categories_Keywords}. Second, we considered the \edit{18} DWMs that included at least one listing in one of the categories in Table~\ref{Categories_Keywords}, thus adding listings to the COVID-19 \textit{mentions} category in our analysis. We investigated the relationship between major COVID-19 events, public attention, and the time evolution of the number of active listings. Third, we quantified the indirect impact that COVID-19 had on all \edit{30} DWMs under consideration by tracking the percentage of listings mentioning the themes of lockdown, delays, and sales. We linked their frequency to major COVID-19 events.

\subsection*{Categories of listings}
\label{sec:results1}

Here, we focus on the \edit{788} COVID-19 specific listings present in our dataset, observed \edit{9,464} times \edit{in the considered time window}. \textit{PPE} is the most represented category, with \edit{355} unique listings (\edit{45.1\%} of COVID-19 specific listings) observed \edit{5,660} times (\edit{59.8\%} of observations of COVID-19 specific listings). The second most represented category is \textit{medicines}, with \edit{228} (\edit{28.9\%}) unique listings observed \edit{1,917} (\edit{20.3\%} of all) times. Some \textit{medicines} listings, which are often sold together, included \edit{38} chloroquine listings, \edit{65} hydroxychloroquine listings, \edit{51} azythromicin listings and \edit{45 Amoxicillin listings. }Other \textit{medicines} included \edit{2} remdesivir listings\edit{, one of the drugs used to treat USA's president Trump~\cite{trump_remdesivir}}. A breakdown of the \textit{medicines} category together with a brief description of the specific drugs can be found in Table~\ref{medicines_table}, and multiple medicines are sometimes sold in the same listing. \edit{Another prominent category was \textit{guides on scamming}, with 99 unique listings (12.6\%). It includes manuals on how to earn money exploiting flaws in COVID-19 related government relief funds, \edit{and others on how} to exploit alleged pandemic related security \edit{weaknesses} (e.g. online banking, delivery systems). A breakdown of the different kinds of guides can be found in table \ref{guides_table}. One DWM (MagBO) was specialised in selling of web domains, like ``coronavirusmasks.in,'' with 50 unique listings (6.3\%).} \edit{Additonally,} we classified \edit{34 (4.3\%)} unique listings as \textit{medical frauds}, which are listings that promised immunity from COVID-19 (no such product exists, at the moment of writing), or supposed devices able to detect COVID-19 in the air. These listings also included illicit drug mixes sold as cures. We also registered \edit{17} test (\edit{2.2\%} of COVID-19 specific listings)\edit{, 3 \textit{fake medical records} (0.4\%)} and 2 ICU \textit{ventilator} (\edit{0.3\%}) listings. More details on these listings together with some examples are reported in Appendix~\ref{Examples_of_listings}. \edit{There were a total of 252 vendors selling COVID-19 specific listings.} 
% The total number of vendors selling COVID-19 specific listings was \edit{252}. 
Additionally, sellers posted multiple \edit{unique} listings. In fact, \edit{88} of them sold \textit{PPE} (\edit{34.9}\%), \edit{106} sold \textit{medicines} (\edit{42.1}\%), \edit{40 sold \textit{guides on scamming} (\edit{15.9\%}), 15 \textit{web domains} (6.0\%), 23} sold \textit{medical frauds} (\edit{9.1}\%), \edit{13} sold \textit{tests} (\edit{5.2}\%), \edit{3 sold \textit{fake medical records} (1.2\%)}, and 2 sold \textit{ventilators} (\edit{0.8}\%). The information in this paragraph is summarized in Table~\ref{Summary_item_found_per_category_listing}.

\begin{table}[h!]
  \centering
\caption{Summary statistics for the considered categories of listings. For each category, we included the number of unique listings, observations, and vendors.
 If the same vendor sold listings in different categories, we counted it as one unique vendor.}
      \begin{tabular}{lccc}
        \hline
        Category & Unique listings & Total observations & Vendors \\ 
        %& Number (\%) & Number (\%) & Number (\%) \\
        \hline
        PPE &  \edit{355 (45.1\%)} & \edit{5,660 (59.8\%)} & \edit{88 (34.9\%)} \\ 
        Medicines & \edit{228 (28.9\%)} & \edit{1,917 (20.3\%)} & \edit{106 (42.1\%)} \\ 
        \edit{Guides on scamming} & \edit{99 (12.6\%)} & \edit{1,244 (13.1\%)} & \edit{40 (15.9\%)} \\ 
        \edit{Web Domains} & \edit{50 (6.3\%)} & \edit{189 (2.0\%)} & \edit{15 (6.0\%)} \\ 
        Medical Frauds & \edit{34 (4.3\%)} & \edit{316 (3.3\%)} & \edit{23 (9.1\%)} \\ 
        Tests &  \edit{17 (2.2\%)} & \edit{51 (0.5\%)} & \edit{13 (5.2\%)} \\ 
       \edit{ Fake Medical Records} &  \edit{3 (0.4\%)} & \edit{9 (0.1\%)} & \edit{3 (1.2\%)} \\
        Ventilators & \edit{2 (0.3\%)} & \edit{78 (0.8\%)} & \edit{2 (0.8\%)} \\ 
        \hline
        COVID-19 & \edit{788 (100\%)} & \edit{9,464 (100\%)} & \edit{252 (100\%)} \\ 
        \hline
      \end{tabular}
\label{Summary_item_found_per_category_listing}
\end{table}

It is important to note that vendors often do not provide complete information on their listings but rather invite direct communication \edit{to facilitate sales}. In \edit{391 (49.6\%)} unique listings, the vendor invited potential customers to communicate via email or messaging applications such as WhatsApp, Wickr Me, and Snapchat. Thus, \edit{511 (64.8\%)} COVID-19 specific listings contained no information about the offered amount of goods, \edit{579 (73.5\%)} did not provide shipping information, and \edit{16 (2.0\%)} did not disclose the listing price.

\begin{figure}[h!]
  \centering
   \includegraphics[width=12cm]{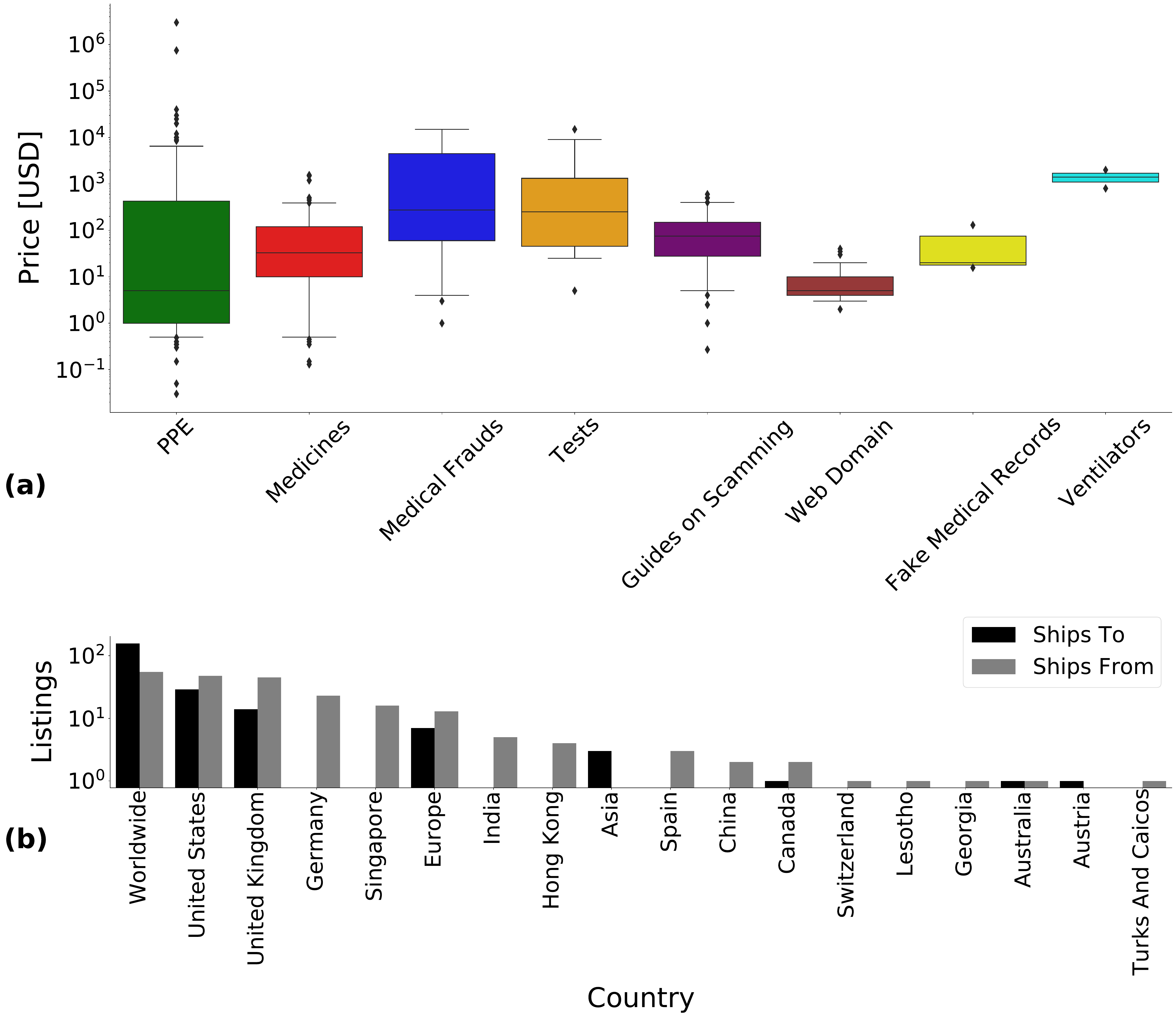}
  \caption{(a) Box plot of the distribution of listing prices for each COVID-19 category. The box ranges from the lower to the upper quartile, with the horizontal line indicating the median. The whiskers extend up to the $5^{th}$ and $95^{th}$ percentiles respectively. The dots represent outliers. (b) Shipping information in COVID-19 specific listings. Note that \edit{545 (or 71.1\%)} of these listings did not provide any shipping information.}
  \label{Price_Quantity_Geography}
\end{figure}

\edit{\textit{PPE} and \textit{web domains} were the least expensive products with a} median price of \edit{5} USD. Followed by \textit{medicines} with \edit{33} USD, \edit{\textit{guides on scamming} with 75 USD, \textit{fake medical records} with 130 USD, }\textit{tests} with \edit{250} USD, \textit{medical frauds} with \edit{275} USD, and \textit{ventilators} with 1,400 USD. The distribution of prices for these categories can be found in Figure~\ref{Price_Quantity_Geography}(a). It shows that many listings had a low price \edit{of} around a few USD or less and only few listings exceeded thousands or more USD. The cumulative value of the detected unique listings was \edit{$563,202$} USD, where we excluded listings with prices larger than $40,000$ USD using manual inspection. When vendors post listings at high price this typically indicates they have halted sales of an item with the expectation of selling it again in the future. We remove these anomalously high priced listings since they would largely overestimate the sales price of actually active listings~\cite{soska2015measuring}. The shipping information declared in the analysed listings involved a total of \edit{18} countries or regions. Most of the vendors are willing to ship worldwide. Shipping from different continents \edit{appears} possible because some vendors explicitly declare \edit{in listing descriptions} that they have multiple warehouses across the globe, while shipping to any continent is done through specialized delivery services. The United States is the second largest exporter and shipping destination. \edit{The United Kingdom} is the third largest exporter \edit{and importer}, \edit{and} no vendors explicitly mentioned \edit{Germany} as a shipping destination \edit{even thought it is the fourth largest exporter}. Complete shipping information is available in Figure~\ref{Price_Quantity_Geography}(b). Some examples of the COVID-19 specific listings are available in the Appendix~\ref{Examples_of_listings_specific}.

  \begin{figure}[h!]
  \centering
   \includegraphics[width=16cm]{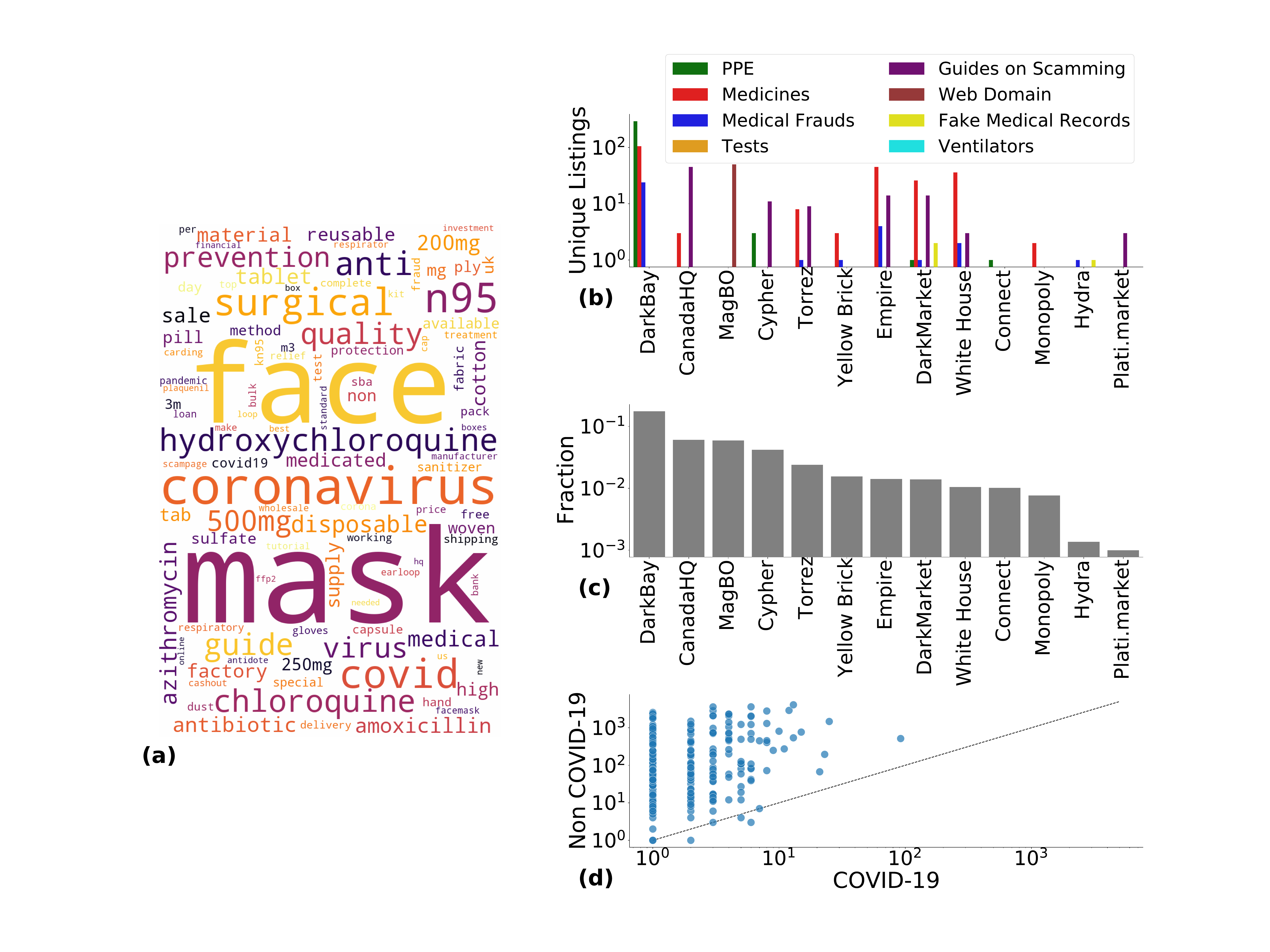}
  \caption{(a) Word cloud for ``Listing title'' in  COVID-19 specific listings. (b) Category breakdown of COVID-19 specific listings in the DWM that offered them. (c) Fraction of vendors selling at least one COVID-19 specific listing. (d) Vendor specialisation. Most vendors responsible for at least one COVID-19 specific listing also sell other listings, and in greater number.}
  \label{Markets_Vendors}
\end{figure}

Figure~\ref{Markets_Vendors}(a) presents a word cloud built from the titles of the selected COVID-19 specific listings. The word cloud was built from 1-grams, meaning single words, excluding common English words and stop words. The COVID-19 pandemic was referred to as either ``coronavirus,'' ``corona,'' ``covid,'' or ``covid19.''. Among COVID-19 \textit{medicines}, ``hydroxychloroquine,'' and ``chloroquine'' were the most popular ones, with fewer mentions of ``azithromycin,'' ``medicated,'' and ``medical'' products in general.

DarkBay/DBay \edit{DWM} contained the majority of the COVID-19 specific listings in our dataset, amounting to \edit{425 (54.0\%)}. The most available unique listings in DarkBay/DBay were \textit{PPE}, which totaled \edit{293}. We also found \edit{105} \textit{medicines}, 24 \textit{medical frauds}, 2 \textit{ventilators}, and 1 \textit{tests}. The number of listings available in the other DWMs was: \edit{94 in Empire (which shut down in August 2020), 50 in MagBO, 49 in DarkMarket, 48 in The Canadian Headquarters, 42 in White House, and 35 in Yellow Brick}. \edit{These numbers together with statistics of the less represented DWMs are as shown in} Table~\ref{BasicStatistics}. The entire breakdown of the number of COVID-19 specific listings detected in each category is available in Figure~\ref{Markets_Vendors}(b).

In Figure~\ref{Markets_Vendors}(c), we ranked the DWMs by their share of vendors selling COVID-19 specific listings. The total number of vendors behind COVID-19 specific listings in our dataset is \edit{252}. Most vendors sold only one COVID-19 specific listing, while few of them sold more than ten different \edit{unique} COVID-19 specific listings. In Appendix~\ref{Supplementary_figures}, we analysed the distribution of COVID-19 specific listings for each vendor. We found that it was heterogeneous according to a power-law with an exponent equal to \edit{$-2.3$} and 80\% of the vendors had fewer than 5 \edit{unique} listings, as shown in Figure~\ref{Appendix_distribution}. \edit{This may imply that vendors of COVID-19 related products have a focus on a specific product category, or are just creating one-off listings to try to make quick money.} In DarkBay/DBay, more than 15\% of the vendors were selling COVID-19 specific listings, while in \edit{MagBO}, The Canadian HeadQuarters, and \edit{Cypher} this fraction was \edit{around 5\%} \edit{(with almost all other DWMs around 1\%)}. \edit{This shows that law enforcement or intelligence intervention should not necessarily be approached evenly across marketplaces but instead focused on select marketplaces first with a higher concentration of COVID-19 specific listings.} Finally, Figure~\ref{Markets_Vendors}(d) shows that essentially no vendors specialised on COVID-19 products, with only \edit{$7$} vendors selling more COVID-19 specific listings than unrelated ones, $4$ of which actually sold just one \edit{or two} COVID-19 specific \edit{listings} overall in our dataset.

\subsection*{Time evolution of DWM listings and public attention}
\label{sec:results2}

\begin{figure}[h!]
  \centering
  \includegraphics[width=12cm]{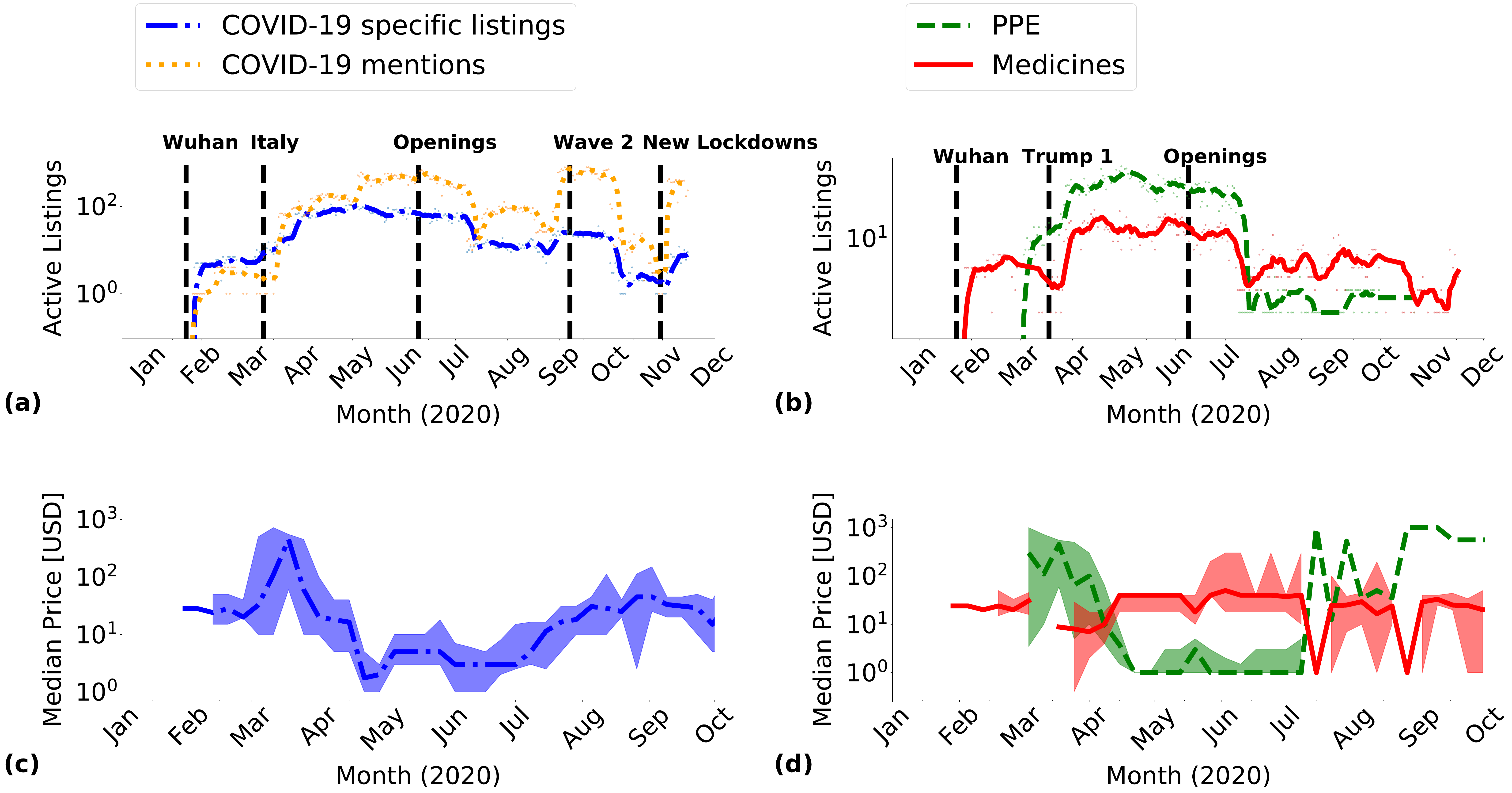}
  \caption{Longitudinal analysis of DWM activity. (a) Seven-days rolling average of active listings mentioning COVID-19 and COVID-19 specific listings. (b) Seven-days rolling average of the observed COVID-19 specific listings in the \textit{medicines} and \textit{PPE} categories. Black dashed vertical lines in panels (a) and (b) corresponded to significant COVID-19 world events, see Appendix~\ref{Timeline_covid}. (c) Seven-days median price with 95\% confidence interval for COVID-19 specific listings. (d) Seven-days median price with 95\% confidence interval for active COVID-19 specific listings in the \textit{PPE} and \textit{medicines} categories.}
  \label{Time_series_Listings_Overall}
\end{figure}

The number of active unique listings evolved over time, as shown in Figure~\ref{Time_series_Listings_Overall}(a). The first COVID-19 specific listing in our dataset appeared on January 28, 2020, following the Wuhan lockdown~\cite{nytimescoronavirus}. In March, lockdowns in many countries~\cite{lockdown_italy,lockdown_uk} corresponded to an increase in the number of these listings, whose number kept increasing until May. In June and July, when worldwide quarantine restrictions started to ease~\cite{lockdown_eased}, we observed a decreasing trend in the selected COVID-19 specific listings, \edit{which continued until November. COVID-19 mentions followed the same trend with two notable exceptions. We observed two sudden increases in COVID-19 in correspondence of the second wave of contagions in Europe in September~\cite{second_wave_europe} and new lockdown measures in November~\cite{new_lockdown_uk}.}
% In correspondence of the second wave of contagions in Europe in September~\cite{second_wave_europe} and new lockdown measures in November~\cite{new_lockdown_uk}, we observed two sudden increases in COVID-19 mentions.} 
Figure~\ref{Time_series_Listings_Overall}(b) shows the evolution of the total number of observed \textit{PPE} and \textit{medicines}, the two most available COVID-19 specific listings in our dataset (see Table~\ref{Summary_item_found_per_category_listing}). \textit{PPE} followed a trend compatible with the overall observations shown in Figure~\ref{Time_series_Listings_Overall}(a), with a peak in May and a \edit{sudden decrease after July, as PPE have gradually become \edit{more} available worldwide with respect to the shortage in the beginning of the pandemic. COVID-19 \textit{medicines} remained approximately stable throughout these months, with a peak after USA president Donald Trump first referred to Chloroquine}~\cite{WhiteHouse_Trump1}. \edit{A different trend was found for COVID-19 \textit{guides on scamming}, which saw spikes in the number of listings in correspondence to event related to relief program measures~\cite{cares_act,sba,heroes_act}. More details can be found in Appendix~\ref{Supplementary_figures}, Figure~\ref{Appendix_guides_on_scamming}.}

The time evolution of the listing prices followed a different pattern. We considered the median price and its 95\% confidence interval of active COVID-19 specific listings in Figure~\ref{Time_series_Listings_Overall}(c), and of active \textit{PPE} and \textit{medicines}, in Figure~\ref{Time_series_Listings_Overall}(d). \edit{Until March, the only COVID-19 specific listings concerned \textit{medicines}, which influenced the overall median price. Afterwards, when \textit{PPE} listings started to appear, they led the variation in the overall median price. In fact, over the entire time window, the median price of \textit{medicines} listings was reasonably stable. \textit{PPE} listings, instead, had a high price for March and most of April}, possibly due to speculation. Interestingly, at the end of April, a vendor named ``optimus,'' active on DarkBay, started selling large quantities of \textit{PPE} at $1$ USD, putting many online listings at the same time, thus drastically reducing the median price, which remained low until July. Overall, ``optimus'' had 91 \textit{PPE} listings during the registered period. \edit{\textit{PPE} median price then increased back to the March level in July, when general worldwide availability of masks for the general population decreased the demand for small quantities of products.} \edit{We report an analysis of the listings price for COVID-19 \textit{guides on scamming} in Figure~\ref{Appendix_guides_on_scamming} of Appendix~\ref{Supplementary_figures}.}

\begin{figure}[h!]
  \centering
 \includegraphics[width=12cm]{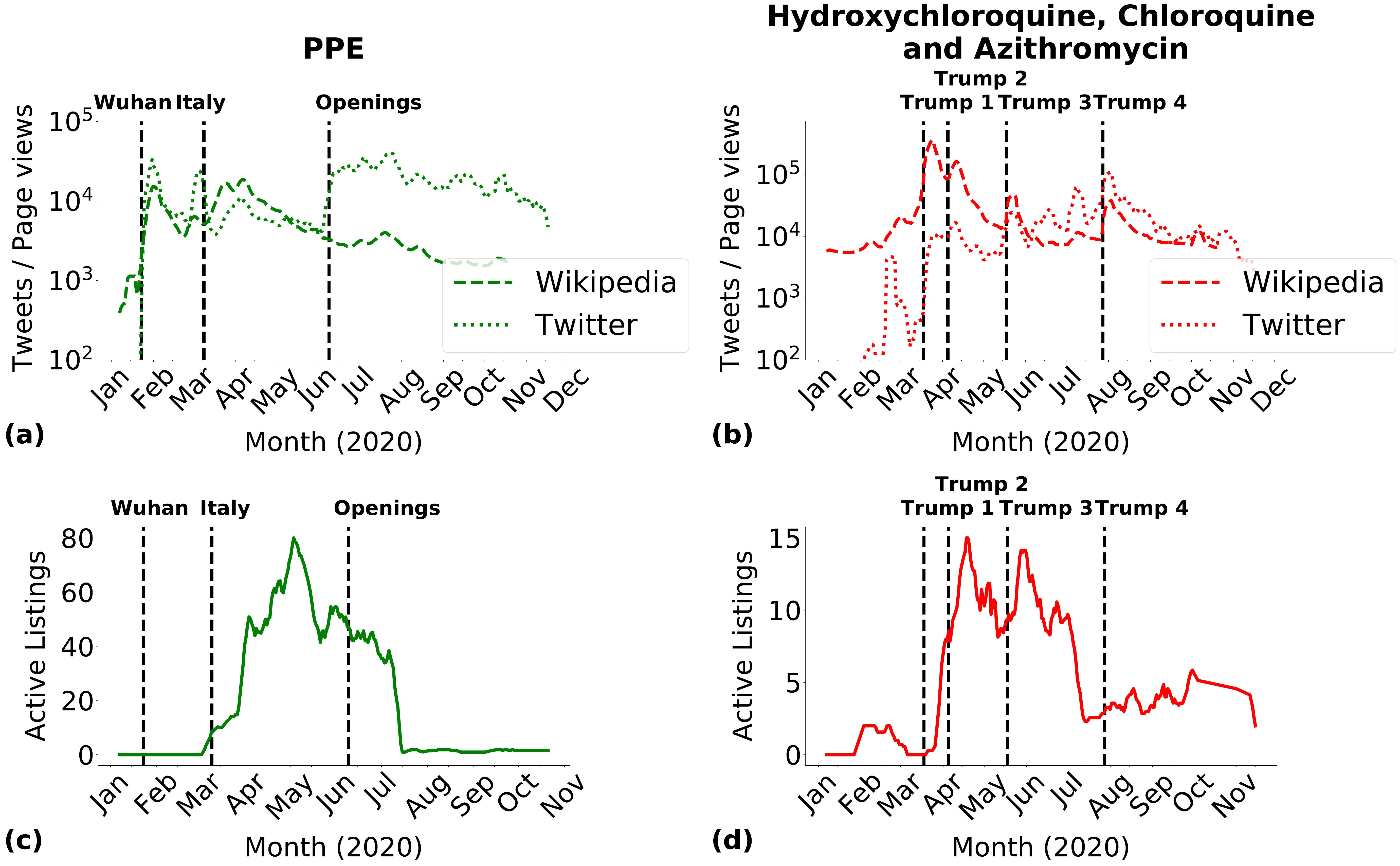}
  \caption{DWMs and public attention. (a)-(c) Seven-days rolling average of active listings selling \textit{PPE}, together with the time evolution of the number of tweets referring to masks and of visits in the relative Wikipedia page visits. (b)-(d) Similar comparison as in panels (a)-(c) but considering active listings of hydroxychloroquine, chloroquine, and azithromycin. Black dashed vertical lines in panels (a) and (b) mark significant events related with COVID-19, see Appendix~\ref{Timeline_covid}. See Appendix~\ref{Supplementary_figures} for panels (a) and (b) with a linear y-axis.}
  \label{Time_series_Twitter_Wikipedia_Listings}
\end{figure}

We also considered tweets and Wikipedia page visits as proxies for public attention\edit{, as already done in prior studies analysing the COVID-19 pandemic}~\cite{gozzi2020collective,gallotti2020assessing,cinelli2020covid}. \edit{We compared trends in public attention with temporal variations in the number of active COVID-19 specific listings on DWMs.} We focused our analysis on the \textit{PPE} category and on relevant \textit{medicines} in our dataset: hydroxychloroquine, chloroquine, and azitrhomycin. Figure~\ref{Time_series_Twitter_Wikipedia_Listings}(a) shows that a first peak in public attention on \textit{PPE} was reached in late January following the Wuhan lockdown~\cite{nytimescoronavirus}. A second peak \edit{occurred} in March~\cite{gallotti2020assessing} when \textit{PPE} listings started to appear in DWMs. The number of \textit{PPE} listings reached their maximum in May. After May, \textit{PPE} listings steadily decreased along with public attention. It is worth noting that May also marked the end of the first wave of contagion in many European countries~\cite{Europe_improving}. \edit{\textit{PPE} listings virtually disappeared in July, as products became more accessible in legal shops. On the contrary Twitter saw a huge spike in June, when many states decided to gradually lift lockdown measures}~\cite{lockdown_eased}\edit{, causing a public debates on mask wearing which increased the twitter signal to stable high levels until November.}

A similar relationship between mass media news, public attention, and DWMs was registered for the listings regarding the three considered \textit{medicines}, as shown in Figures~\ref{Time_series_Twitter_Wikipedia_Listings}(b) and (d). \edit{Four} peaks in public attention were detected after four declarations from President Trump about these \textit{medicines}~\cite{WhiteHouse_Trump1,Trump_2,Trump_3, Trump_4}. The number of active \textit{medicines} listings closely followed. However, a closer look reveals the different shapes of the Wikipedia page visits, tweets, and DWMs curves. Wikipedia saw a very high peak of page visits after the first declaration from President Trump~\cite{WhiteHouse_Trump1}, and smaller peaks in correspondence in the following declarations. Tweets instead saw peaks of attention of increasing height. DWM listings on the contrary were much steadier in time and with little variation in the number of active listings \edit{throughout the first wave of the pandemic, while decreasing to a lower steady availability from the summer.}

\subsection*{Impact of COVID-19 on other listings}

\label{sec:results3}
We considered the indirect impact of COVID-19 on all the \edit{30} DWMs in our dataset. We analyzed all listings in these DWMs (COVID-19 related and beyond), and looked at listings mentioning: lockdown, using keywords ``lockdown'' or ``quarantine,'' delay, using ``delay'' or ``shipping problem,'' and sales, using ``sale,'' ''discount,'' or ``special offer.'' Examples of listings reporting these keywords are available in Appendix~\ref{Examples_of_mentions}.

Figure~\ref{Time_series_Keywords}(a)-(b)-(c) shows the percentage of all listings mentioning these themes over time. The percentage of all listings in the \edit{30} DWMs mentioning lockdown never exceeded 1\%, as illustrated in Figure~\ref{Time_series_Keywords}(a). It reached its maximum in \edit{November, when Europe started new lockdown measures}~\cite{new_lockdown_uk}. \edit{Other peaks occurred in April and September, when nations first started to implement these measures~\cite{nytimescoronavirus, lockdown_italy,lockdown_uk} and at the beginning of the second wave of contagions in Europe~\cite{second_wave_europe}, respectively.} Delay mentions reached local peaks in March and May. These peaks occurred after major COVID-19 events, such as lockdowns~\cite{lockdown_italy,lockdown_uk} and the situation in Europe starting to improve~\cite{Europe_improving}, respectively. \edit{Two global peaks, instead, were reached in September and November, when cases started to surge again in Europe~\cite{second_wave_europe} and when Europe started new lockdown measures~\cite{new_lockdown_uk}},
as shown in Figure~\ref{Time_series_Keywords}(b). A similar pattern was visible for the percentage of all listings mentioning sales. In addition, we observed that sales had a first peak corresponding to the New Year, which is a common practice of many \edit{offline regulated} shops, as displayed in Figure~\ref{Time_series_Keywords}(c). Despite observing that the increase in the percentage of all listings mentioning sales, delays, and lockdown followed major events related to the pandemic, not all of these listings also mentioned COVID-19. We further researched this by plotting which percentage of the relative listings also mentioned COVID-19 in Figure~\ref{Time_series_Keywords}(d). The percentage of listings mentioning that current sales were due to COVID-19 was less than 1\%, while mentions of delays reached up to 40\%. For lockdown it was 100\%, as one can expect since lockdowns exist because of COVID-19. In the three selected cases, the percentages of listings mentioning COVID-19 \edit{followed} the global awareness about the current pandemic: \edit{increasing trends from January to the July~\cite{nytimescoronavirus,lockdown_italy,lockdown_uk, 1M_ww_cases}, less attention during the summer~\cite{Europe_improving}, and a returning increase in September and November~\cite{second_wave_europe, new_lockdown_uk}.}

  \begin{figure}[h!]
  \centering
  \includegraphics[width=12cm]{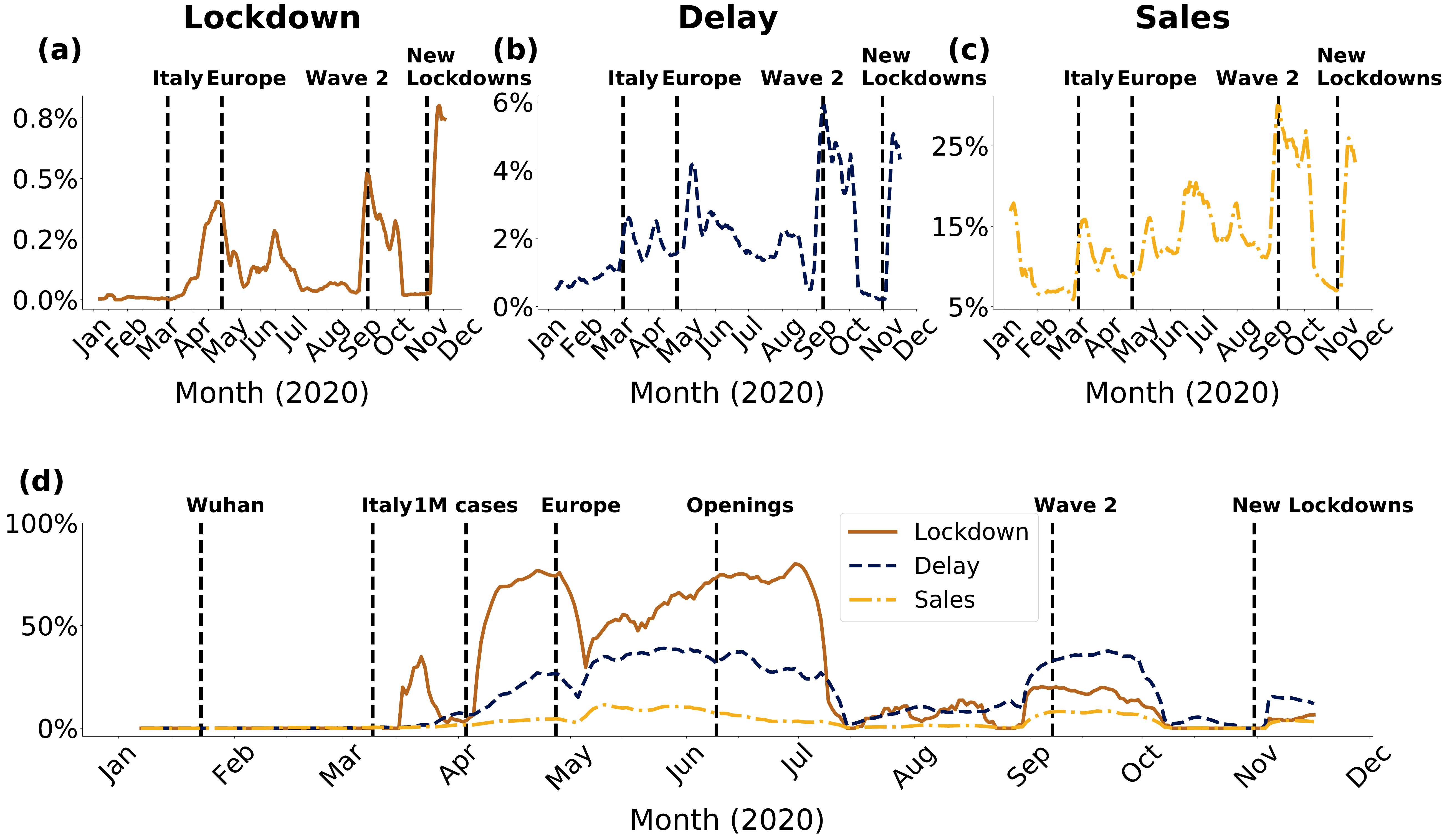}
  \caption{Percentage of all active listings mentioning the themes lockdown, delay and sales in panels (a), (b), (c), respectively. (d) Percentage of active listings in panels (a), (b), (c) that mentioned also COVID-19 in their listings. Black dashed vertical lines in panels (a), (b), and (c) corresponded to major COVID-19 events, see Appendix~\ref{Timeline_covid}.}
  \label{Time_series_Keywords}
\end{figure}

\section*{\edit{Discussion}}

We investigated the presence of listings related to COVID-19 in \edit{30} DWMs, monitored over a \edit{ten}-months period in 2020. We considered COVID-19 specific listings and COVID-19 \textit{mentions}, \edit{found} them in \edit{13} and \edit{18} DWMs, respectively. COVID-19 specific listings totaled \edit{788 unique products} and represented less than 1\% of our dataset. 
The majority of COVID-19 specific listings offered \textit{PPE} (\edit{45.1}\%), followed by \textit{medicines} (\edit{28.9}\%), \edit{\textit{guides on scamming} (12.6\%), \textit{web domains} (6.3\%),} \textit{medical frauds} (\edit{4.3}\%), \textit{tests} (\edit{2.2}\%), \edit{\textit{fake medical records} (0.4\%)} and \textit{ventilators} (\edit{0.3}\%). Most COVID-19 specific listings did not report the quantity sold (\edit{64.8}\%) or shipping information (\edit{73.5}\%). Almost half of these listings invited potential customers to communicate via email or messaging applications, like WhatsApp (\edit{49.6}\%). \edit{Although} direct communication fosters a trustworthy vendor-buyer relationship and may lay the ground for future \edit{transactions} outside DWMs\edit{, it also} exposes users to higher risk of being traced by law enforcement~\cite{arun2019whatsapp}.

In our dataset, DarkBay/DBay is featured prominently among DWMs offering COVID-19 specific listings. Ranking in the top 100 sites in the entire dark web~\cite{OnionLive:DarkBay}, DarkBay/DBay is regarded as the eBay of the dark web because it offers more listings categories than other DWMs~\cite{DNStatsDarkBay}. It was also frequently accessible during the period of time monitored during this research, with an uptime of \edit{80}\%, higher from the 77\% uptime of Empire, the largest global DWM at the time of writing~\cite{Markets:uptime}.

Our work corroborates previous findings and expands them in several ways. To the best of our knowledge, the most extensive report to date examined the presence of COVID-19 specific listings in 20 DWMs on one single day (April 3, 2020)~\cite{broadhurstavailability}. Despite only a subset of overlapping DWMs between that report and our study, (\edit{Cypher}, DarkBay/DBay, DarkMarket, Empire, \edit{Monopoly}, Venus Anonymous, White House, and Yellow Brick) we both assessed that COVID-19 specific listings constituted less than 1\% of the total listings in the DWMs ecosystem. These listings were mostly \textit{PPE}, followed by \textit{medicines} and they were found in only a few DWMs, while non COVID-19 specific listings were widespread. 
 
An important novelty of the present study is the analysis of the temporal evolution of DWM behaviour and its relationship to public attention, as quantified through tweets and Wikipedia page visits. Following the Wuhan lockdown~\cite{nytimescoronavirus}, we observed a first peak in public attention~\cite{cinelli2020covid}, and a corresponding emergence of the COVID-19 specific listings. A second peak in public attention occurred in March, when quarantine measures were adopted by many European countries~\cite{lockdown_italy,lockdown_uk}. Again, during the same period, the number of COVID-19 specific listings sharply increased. When worldwide quarantine began to ease~\cite{lockdown_eased} in many countries, in June and July, we registered a decrease in public attention and in available COVID-19 specific listings. \edit{Towards the end of 2020, we did not detect significant variations in COVID-19 specific listings and public attention, in correspondence of the second wave of contagions~\cite{second_wave_europe} and new lockdown measures in Europe~\cite{new_lockdown_uk}. Both vendors of COVID-19 specific listings and public attention have adapted to the COVID-19 pandemic and react more smoothly to its development.}

Listing prices correlated with both variations in public attention and individual choices of a few vendors. Median price experienced a sharp increase in March, probably due to speculation, and then decreased in April due to the choice of a single vendor responsible for 91 listings, named ``optimus.'' The vendor sold a large quantity of \textit{PPE} at $1$ USD only, which constituted the 37\% of active \textit{PPE} listings in April. %We observed a total of 450 unique listings (mostly recreational drugs) sold by the vendor from March to July 2020. 
Finally, we observed an increase in the percentage of all listings citing delays in shipping and sale offers, which peaked in March, May\edit{, September, and November}. Similar to a prior work that found Wikipedia page visits of a given drug to be a good predictor for its demand in DWMs~\cite{miller2020predicting}, we provide further evidence that the DWMs ecosystem is embedded in our society and \edit{responds in line with} social changes~\cite{bancroft2019darknet}. The DWMs ecosystem swiftly reacted to the pandemic by offering goods in high demand, and even offering vaccines already in March, when no tested vaccination existed.

Our research shares some limitations with previous studies, namely that not all active DWMs were surveyed. For instance, we did not analyse \edit{12} of the DWMs explored in the previous report~\cite{broadhurstavailability}. It must be noted, however, that the number of active DWMs is constantly changing due to closures or new openings~\cite{elbahrawy2019collective} and obtaining full coverage is challenging due to the active efforts of DMWs to obstruct research studies and law enforcement investigations, for example through the use of CAPTCHAs. \edit{Another limitation is the lack of reliable fully automated annotation method: this forced us to manually annotate listings and thus limited our analysis to listings only directly related to COVID-19. One key problem to be solved in this regard is the presence of false positives when doing a keyword search. In the presence of a more automated pipeline, one could extend this analysis to a more precise quantification of the effect of the pandemic on traditional DWMs' goods like weapons, drugs or digital goods.}

%Also, the DWM ecosystem had made an extensive use of CAPTCHAs to purposely obstruct research studies and law enforcement investigations. In spite of these limitations, our study is currently the most extensive and accurate analysis about the effects that COVID-19 was having over time on illicit online trade. We are presently the only ones to monitor the availability of COVID-19 specific listings since the beginning of the outbreak. 

\section*{\edit{Conclusion}}

By revealing that DWMs listings of goods related to COVID-19 exist and that they are correlated with public attention, we highlight the need for a close monitoring of the online shadow economy in the future months, in order to control and anticipate dangerous effects of the COVID-19 infodemic~\cite{cinelli2020covid, gallotti2020assessing}. 
% For example, we expect that initial delays in the availability of a cure and/or vaccine would dramatically increase public interest in the online shadow economy, posing concrete risks to public health.
We plan to improve our analysis of DWM activity by increasing the number of monitored DWMs and conducting a more extensive analysis of the impact on the pandemic on overall DWM trade by considering changes in prices of non-COVID-19 specific listings, such as drugs, weapons or malware. \edit{We released a new website~\cite{monitoring_effort}, where we will provide constant updates on the effect of the pandemic on DWMs.} 

We anticipate that our work \edit{will interest a wide audience and spark new research. Future research work may further explore the behaviour of DWMs over time, potentially extending the spectrum of monitored goods and relating the observed trends to specific social changes. Policy makers and public agencies (especially those focused on protecting consumer rights and health) can use our findings to better assess and shape the effects of legislation on the shadow economy~\cite{COVID19Drugs}. Practitioners may gain insights on how DWMs posit additional threats to public health. Uninformed citizens exposed to waves of misinformation, such as the ones related to hydroxychloroquine, chloroquine, and azitrhomycin \cite{WhiteHouse_Trump1, Trump_2, Trump_3, Trump_4}, may be tempted to shop on DWMs thus exposing themselves to serious health risks. Moreover, the availability of regulated products currently in shortage in the traditional economy undermines anti-price gouging regulations and licit businesses which sell the same products.}

%\nocite{oreg,schn,pond,smith,marg,hunn,advi,koha,mouse}

%%%%%%%%%%%%%%%%%%%%%%%%%%%%%%%%%%%%%%%%%%%%%%
%%                                          %%
%% Backmatter begins here                   %%
%%                                          %%
%%%%%%%%%%%%%%%%%%%%%%%%%%%%%%%%%%%%%%%%%%%%%%

%\begin{backmatter}

\section*{Competing interests}
  The authors declare that they have no competing interests.

\section*{Author's contributions}

ABa, AT and AG conceived of the project. ABa coordinated the project. All authors designed the research. MA and IG provided the data. ABr, MN, MA and IG preprocessed and analysed the data. All authors analysed the results and wrote the manuscript. All authors approved the final version of the manuscript.

\section*{Acknowledgements}
A.Br., M.N., A.T., A.G and A.Ba. were supported by ESRC as part of UK Research and Innovation’s rapid response to COVID-19, through grant ES/V00400X/1. M.A and D.M., acknowledge support from the U.S. National Science Foundation grant 1717062.

\section*{Correspondence}
Correspondence and requests for materials should be addressed to \\Andrea Baronchelli: Andrea.Baronchelli.1@city.ac.uk.

\bibliographystyle{unsrt}
%\bibliography{bibliography}

\clearpage
\onehalfspacing
\begin{Large}{\begin{center}
{\bf{The COVID-19 online shadow economy}}
\end{center} 
 }
\end{Large}
\thispagestyle{empty}
\tableofcontents
\clearpage

\appendix

\section{Data pre-processing}
\label{Data_preprocessing}

In the following, we describe the DWMs dataset in more details, by focusing on how listings were stored and how we formed the COVID-19 categories in Table~\ref{Categories_Keywords}, that is, \textit{PPE}, \textit{medicines}, \edit{\textit{guides on scamming}, \textit{web domains},} \textit{medical frauds},  \textit{tests}, \edit{\textit{fake medical records}}, \textit{ventilators}, and COVID-19 \textit{mentions}.

\begin{table}[ht]
  \centering
\caption{Selected attributes of the listings under consideration along with a brief explanation of their respective purposes.}
      \begin{tabular}{ll}
        \hline
        Attribute of a listing & Explanation \\ \hline
        ``Listing body'' & Description of the listing as it appears in the DWM \\
        ``Listing title'' & Title of the listing  as it appears in the DWM \\
        ``Marketplace name'' & Name of the DWM \\
        ``Shipping information'' & Where the listing is declared to ship from and to \\
        ``Time'' & When the listing is observed \\
        ``Quantity'' & Quantity of the listing sold \\
        ``Price'' & Price of listing \\
        ``Vendor'' & Unique identifier of the vendor \\
        \hline
      \end{tabular}
\label{Listing_attributes}
\end{table}

The listings appearing on the DWMs were crawled and stored according to selected attributes. While a brief explanation of these attributes is already presented in Table~\ref{Listing_attributes}, here we focus on those attributes which involved some pre-processing before the analysis, that is, ``Shipping information,'' ``Quantity,'' and ``Price.'' The ``Shipping information'' attribute was initially stored considering what the vendor declared. Then, it was standardised among vendors to correct any misspellings, using the standard python library \textit{pycountry}. Vendors may declare a specific country, like United States, a continent, like Europe, or the entire world, which we standardise here as worldwide. The ``Quantity'' attribute was instead retrieved from the title of the listing using Facebook open-source library \textit{Duckling}~\cite{duckling}, then it was manually checked and corrected during an annotation process. The ``Price''  attribute on DWMs was displayed in the listings in various currencies, such as cryptocurrencies and fiat currencies. In order to standardise and properly compare listing prices, we converted prices to USD at the daily conversion rate. Rates were taken from Cryptocompare~\cite{cryptocompare} for cryptocurrencies, and from the European Central Bank~\cite{european_central_bank} for fiat currencies.

The attributes ``Listing body'' and ``Listing title'' in Table~\ref{Listing_attributes}, representing the title and description of the listings, were used to select the COVID-19 categories in Table~\ref{Categories_Keywords}. To this end, we prepared two sets of keywords as shown in
Table~\ref{Keywords}. Every selected COVID-19 listing contained either a word in the ``Listing body'' that matched one keyword in the first set or a word in the ``Listing title'' that matched one keyword in the second set. The rationale behind this choice was that the listing title was usually more precise on the product sold, whereas the body might contain promotions of other items the vendor was selling in other listings. At the same time, the vendor might mention COVID-19 in the body for various reasons, which we analysed in the main text. In order to classify listings in either COVID-19 specific listings (that is,  \textit{PPE}, \textit{medicines}, \edit{\textit{guides on scamming}, \textit{web domains},} \textit{medical frauds},  \textit{tests}, \edit{\textit{fake medical records}}, \textit{ventilators}) or COVID-19 \textit{mentions}, we ran a regex query in google \emph{bigquery}. We remark that the chosen method returned words containing a string equal to one of our keywords. For instance, with the keyword chloroquin, we detected also chloroquine and hydroxychloroquine. After this automatic filtering step, we manually checked the selected COVID-19 related listings to further improve the accuracy of our sample. In order to minimize human error, at least two authors of the present manuscript checked each of these listings. A limitation of our approach was that keywords considered were in English. Therefore, even if drug names such as chloroquine were common to many languages and we detected some listings in a non-English language, our sample of COVID-19 related listings was biased toward the English language.

\begin{table}[ht]
  \centering
\caption{Keywords used to sample COVID-19 specific listings from the DWMs in Table~\ref{BasicStatistics}.}
      \begin{tabular}{l}
        \hline
          First set of keywords checked against the words included in the attribute ``Listing body'' in Table~\ref{Listing_attributes} \\ \hline
        corona virus, coronavirus, covid, covid-19, covid19         \\ \\ \hline
        Second set of keywords checked against the words included in the attribute ``Listing title'' in Table~\ref{Listing_attributes} \\ \hline
        anakinra, antidote, antiviral, azithromycin,
        baloxavir, baricitinib, bemcentinib, chloroquin, \\
        corona virus, coronavirus, covid, covid-19, covid19,
        darunavir, dexamethason, diagnosis,  \\
        diagnostic, favipiravir, ganciclovir, glove, gown,
        lopinavir, marboxil, mask, n95, n99, \\
        oseltamivir, prevention, remdesivir, repurposed,
        ribavirin, ritonavir, sanitiser, \\
        sanitizer, sarilumab, siltuximab, surgical, 
        thermo scanner, thermo-scanner, \\
        thermometr, thermoscanner, tocilizumab, umifenovir, vaccine,
        ventilator, \edit{ciprofloxacin,} \\ 
        \edit{doxyciclin, metronidazol, amoxicillin} \\
         \hline
      \end{tabular}
\label{Keywords}
\end{table}

While each listing had an associated url to determine its uniqueness, which allowed us to track listing over time, vendors receiving bad reviews sometimes put identical copies of the same listing online. To overcome this issue and correctly count the number of listings, we determined a listing as unique if it had the combination of ``Listing body,'' ``Listing title,'' ``Marketplace name,'' and ``Vendor'' different than any other listings. For instance, if two listings had the same title and body but were sold in two different DWMs, we considered them as two different listings. Also, we considered at most one observation for each unique listing per day. The total number of unique listings and observations of these listings in each DWM is available in Table~\ref{BasicStatistics}.

\section{Examples of listings related with COVID-19 in dark web marketplaces}
\label{Examples_of_listings}

Here, we present detailed examples of the selected listings. We consider both COVID-19 specific listings and COVID-19 \emph{mentions}.   

\subsection{COVID-19 specific listings}
\label{Examples_of_listings_specific}

The most popular category of COVID-19 specific listings was \textit{PPE}, which included mainly face masks. We detected listings selling small quantities of masks, like ``KN95 Face Mask for Corona Virus box of 50'' priced at $50$ USD, while others proposed wholesale deals, as in ``AFFORDABLE 20 BOXS OF SURGICAL FACE MASK (WHOLESALE PRICE)'' in which 5000 masks were available at $2,000$ USD.

The second most popular COVID-19 category was \edit{\textit{guides on scamming} and includes listings explaining how to stole several kinds of COVID-19 related relief funds. Specifically, a subset of these listings were about the Small Business Administration loan in the USA. They provided step-by-step instructions, with constant updates to ensure the scam activities were effective. One listing in particular suggested: ``I do not recommend taking more than 10,000 of the approved amount, because after that cashing out becomes a little harder.'' The price of this listing was 113 USD.}

\edit{MagBo was the only DWM selling listings in the \textit{web domains} category. These listings may cause a potential threat to public health. They may be the actor of several phishing activities or sell scams. Examples of these web domains were ``coronavintheworld.com,'' ``covid-conspiracy.net,'' and ``coronavirusmasks.in.'' Prices of these domains were low and less than 10 USD.}

\edit{Listings on the} \textit{medicines} \edit{category were} composed mostly by chloroquine, hydroxychloroquine, and azythromicin. We registered several wholesale deals, as in ``9000 tabs hydroxychloroquine 200mg (USA AND CANADA ONLY)`` where 9,000 tabs were sold for 1,194 USD. The smallest quantity we detected was 50 pills ``chloroquine 50pills for 250\$,'' sold at 250 USD. We also noticed that vendors often specified the size of the pill, being it 200mg, 250mg, or 500mg. The azythromicin was usually sold together with hydroxychloroquine as a prescription against COVID-19. One example of it was ``hydroxychloroquine sulfate 200mg and azithromycin 250mg,'' where an unknown quantity of these drugs was sold for $40$ USD. 

In the COVID-19 category of \textit{medical frauds}, the most prominent listings were vaccines. Despite at the moment of writing of \edit{the first version of this} manuscript (July 2020), vaccines are far from being actually developed, they were sold in DWMs since March. These listings included both low price vaccines like ``complete order free shipment COVID19 VACCINE,'' sold at just $200$ USD, or high price one like ``Covid-19 Vaccine. Lets keep it low key for now,'' priced at $15,000$ USD. In addition, among the listings in the \textit{medical frauds} category, one could find potentially dangerous illicit drug mixes with claimed curative power against COVID-19, like ``Protect yourself from the corona virus:'' a marijuana based drug mix supposedly helpful in recovery from coronavirus infection. Other \textit{medical frauds} included a $300$ USD ``CORONAVIRUS DETECtOR DEVICE, SAVE LIVES NOW'' or a $1,000$ USD ``Buy CORONAVIRUS THERMO METER.''

\textit{Tests} category of COVID-19 specific listings count a few different items. We detected listings in the \textit{tests} category both at low quantities, such as, ``25 pcs COVID-19 (coronavirus) quick test,'' sold for $430$ USD, or at very large one, like ``Corona Virus Test / Covid-19 Test Kits ( 5000Pcs),'' for a price of 7,500 USD. 

\edit{The three listings in the \textit{fake medical records} categories can be used to fake COVID-19 diagnosis. One of these listings said in its title: ``Novelty/Fake Medical Records! Any diagnosis, custom made.'' And in its body claimed ``The right medical excuse can get you out of anything, and open many doors,'' with a following disclaimer ``IT IS UP TO YOU TO USE THESE ETHICALLY AND LEGALLY!'' The price for this listing was 20 USD only, which could favour its wide adoption.}

The two listings in the \textit{ventilator} category were ICU ventilators. They were advertising fundamental hospital instrument, such as, ``ICU Respiratory Ventilators , Emergency Room Vents'' sold at $800$ USD or ``BiPAP oxygen concentrator ventilato Amid Covid-19'' for $2,000$ USD.

\subsection{COVID-19 mentions}
\label{Examples_of_mentions}

We describe three examples of listings in the COVID-19 \textit{mentions} category. The listing with title ``Best Organic Virginia Bright Tobacco Premium quality 600g'' refers to the lockdown in its body as ``unfortunately we have to respect coronavirus lockdowns, in order to ensure as much security as possible, we had to choose one type of shiping that is unfortunately much more expensive while lockdowns last.'' Another listing with title ``(Out of Stock! Lower Price for Pre-orders Only) Testosterone Enanthate 250mg/ml - 10ml - Buy 4 Get 1,'' mentions in the body that they ``are currently out of stock of this product due to our oil suppliers not being able to get their raw powders shipped to them because of the Coronavirus'' and they ``have lowered the price a little to help make up for this delay.'' A third listing mention a sale directly in the title ``COVID-19 SPECIAL OFFER 1GR CROWN BOLIVIAN COCAINE 90\% £65,'' and link the discount with the distress caused by the pandemic.

\section{Timeline of the COVID-19 pandemic}
\label{Timeline_covid}

In this Section we aim at providing a summary of the main events related to the pandemic, focusing on the ones cited in the main text and listed in Table~\ref{COVID-19_dates}. This is by no means a complete summary of the COVID-19 pandemic timeline.
 
The first event to gain international attention and make the public aware of the coronavirus was the decision from China to lockdown the city of Wuhan, first epicenter of the pandemic, on January 23, 2020~\cite{nytimescoronavirus}. The virus then found its way to Europe, where the first country to be heavily hit by the pandemic was Italy. The Italian government decided to lockdown the entire country on March 9, 2020~\cite{lockdown_italy}. The virus rapidly spread in Europe and internationally, with cases appearing more and more in the United States, leading USA's President Donald Trump to first take a stance on the possibility of using chloroquine to cure individuals infected from COVID on the March 18, 2020~\cite{WhiteHouse_Trump1}. The epidemic started to heavily hit the United States and cases were surging almost everywhere in the world: 70 days after the lockdown of Wuhan, the worldwide count of infections had already surpassed 1 Million cases on April 3, 2020~\cite{1M_ww_cases}. \edit{On March 27, President Trump signed the Cares Act with the first economic aids to whose affected by COVID-19~\cite{cares_act}. After that, he }explicitly promoted the use of hydroxychloroquine on April 5, 2020, before any official medical trial ended~\cite{Trump_2}. In April the situation started to become asymmetric. In Europe, thanks to the many policies in place, the COVID-19 became less threatening~\cite{Europe_improving} and lockdowns started to be eased~\cite{lockdown_eased}. \edit{USA and other countries were instead seeing a rise in cases, and the USA Senate prolonged the small business rescue fund~\cite{sba}}.

\edit{In May, President Trump declared he was now taking Hydroxychloroquine preventively against COVID-19~\cite{Trump_3}, while in July, he posted a video (labet banned by Twitter) diffusing misinformation about the medicine~\cite{Trump_4}. The second wave of contagions hit Spain in September, and few weeks later the entire Europe~\cite{second_wave_europe}\edit{, while The USA saw the failing on the negotiations around a second relief package~\cite{heroes_act}}. Several new lockdown measures have took place in November in Europe~\cite{new_lockdown_uk} and, through that month, the number of COVID-19 related new infections has started to reduce. In the meantime, the USA is continuing to register a high number of new contagions.}

\begin{table}[ht]
  \centering
\caption{Significant COVID-19  events. We defined an acronym for each event and reported it in the main text plots. Please note that this list does not intend to be exhaustive or to establish a ranking between events.}
      \begin{tabular}{llc}
        \hline
          Date & Event & Acronym \\ \hline
2020-1-23  & Wuhan Lockdown~\cite{nytimescoronavirus} & Wuhan \\
2020-3-9 & Italy Lockdown~\cite{lockdown_italy} & Italy\\
2020-3-18 & USA's President Trump first refers to chloroquine~\cite{WhiteHouse_Trump1} & Trump 1 \\
2020-3-27 & USA's President Trump signs the CARES act~\cite{cares_act} & \edit{Cares} \\
2020-4-3 & 1M COVID-19 cases worldwide~\cite{1M_ww_cases} & 1M cases \\
\multirow{2}{*}{2020-4-5} & USA's President Trump promotes the use of & \multirow{2}{*}{Trump 2} \\
 & chloroquine and hydroxychloroquine against COVID-19~\cite{Trump_2} & \\
2020-4-24 & COVID-19 cases in Europe are beggining to slow down~\cite{Europe_improving} & Europe \\
\multirow{2}{*}{2020-5-18} & USA's President Trump declares he is taking & \multirow{2}{*}{Trump 3}  \\
 & hydroxychloroquine preventively against COVID-19~\cite{Trump_3} &  \\
2020-6-9 & Governments start to lift lockdown measures around the world~\cite{lockdown_eased} & \edit{Openings} \\
2020-6-30 & USA senate agrees to extend small business rescue~\cite{sba} & \edit{SBA} \\
\multirow{2}{*}{2020-7-28} & Twitter limits Donald Trump Jr's account & \multirow{2}{*}{\edit{Trump 4}} \\
 & for posting COVID-19 misinformation~\cite{Trump_4} & \\
 \multirow{2}{*}{2020-9-7} & Spain is the first country in Europe & \multirow{2}{*}{\edit{Second wave}} \\
 & to record half a million COVID-19 cases~\cite{second_wave_europe}  &  \\
 2020-9-10 & Negotiations for the Heroes act keep failing~\cite{heroes_act} & \edit{Heroes} \\
2020-10-31 & PM announces four-week England lockdown~\cite{new_lockdown_uk} & \edit{New lockdowns} \\
      \end{tabular}
\label{COVID-19_dates}
\end{table}

\section{Supplementary material}
\label{Supplementary_figures}

In this Section we provide additional material that support our main findings. In Table~\ref{market_details} we provide more details on the \edit{30} DWMs considered in our study. In particular we 
indicate the main specialization of the DWMs, i.e., the main category of products sold. If it is ``Mixed'', it means that the DWM is not specialised in any particular category of goods. In the description we instead put information on the DWMs, with more details where available. All this information has been researched and compiled by the authors, with particular help given by Flashpoint Intelligence~\cite{flashpoint}.

In Table~\ref{medicines_table} we provide a Table reporting the different COVID-19 related medicines which were found in the listings. The medicines were selected as they have been found or claimed to be effective against COVID-19~\cite{covid19_medicines}.
The number of listings related to each medicine is also reported, noting that some listings sell more than one medicine (e.g. listings selling both hydroxychloroquine and azitrhomycin). 

\begin{figure}[h!]
  \centering
  \includegraphics[width=12cm]{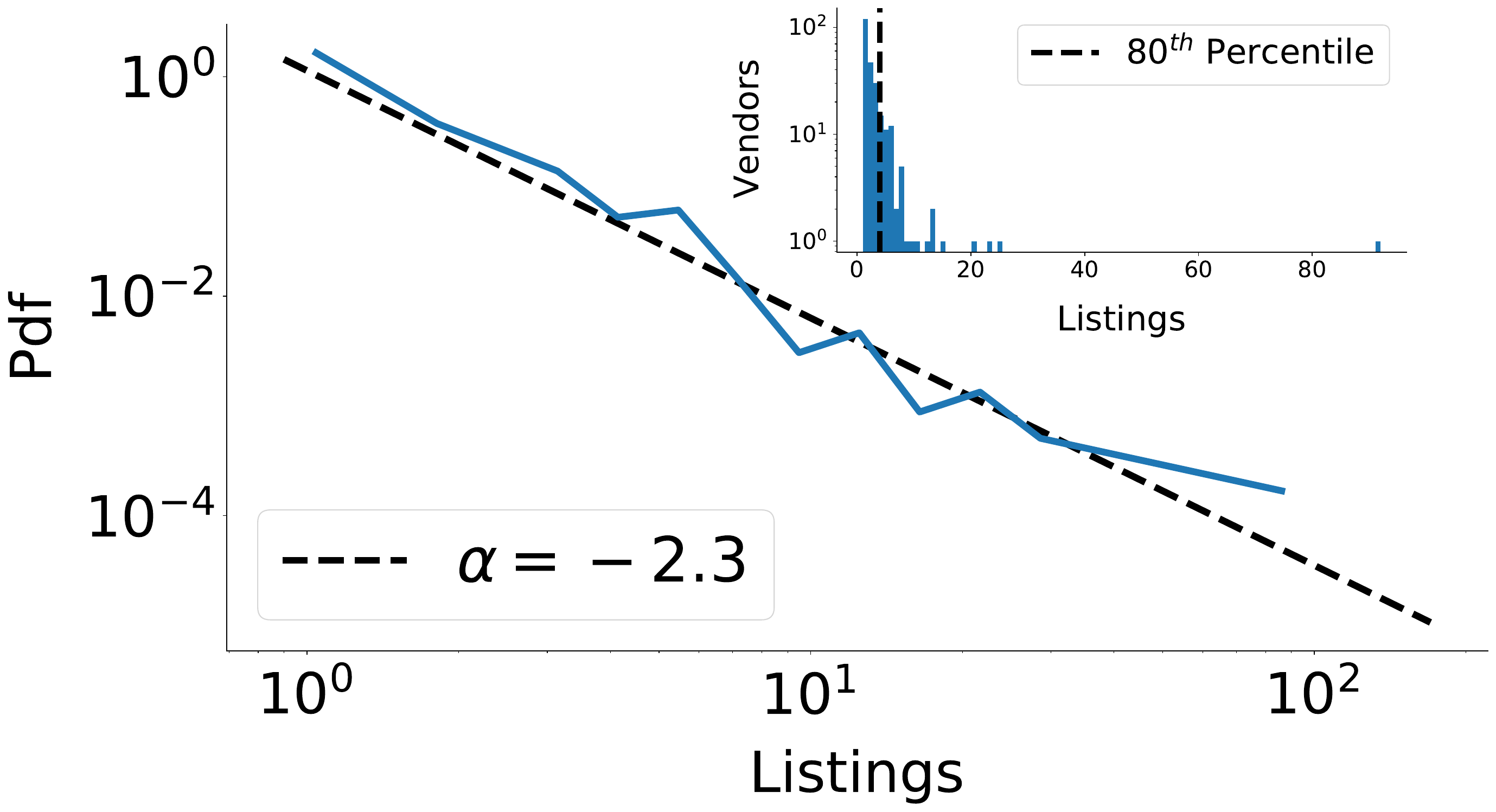}
  \caption{ Probability distribution function (Pdf) for the number of listings per vendor. The power law fit results in an exponent of \edit{-2.3}. In inset, the histogram of the number of listings per vendor, with a vertical line showing the $80^{th}$ percentile.}
  \label{Appendix_distribution}
\end{figure}

In Figure~\ref{Appendix_distribution} we plot the distribution of listings per vendor in log-log plot, showing a clear power-law shape with exponent -2.0. In the inset of Figure~\ref{Appendix_distribution}, we show the histogram using linear spacing, through which we understand that most vendors sold very few COVID-19 specific listings, while few vendors going as high as 91 different listings. We noted that 80\% of the vendors had indeed less or equal than 5 listings. 

\begin{figure}[ht]
  \centering
  \includegraphics[width=12cm]{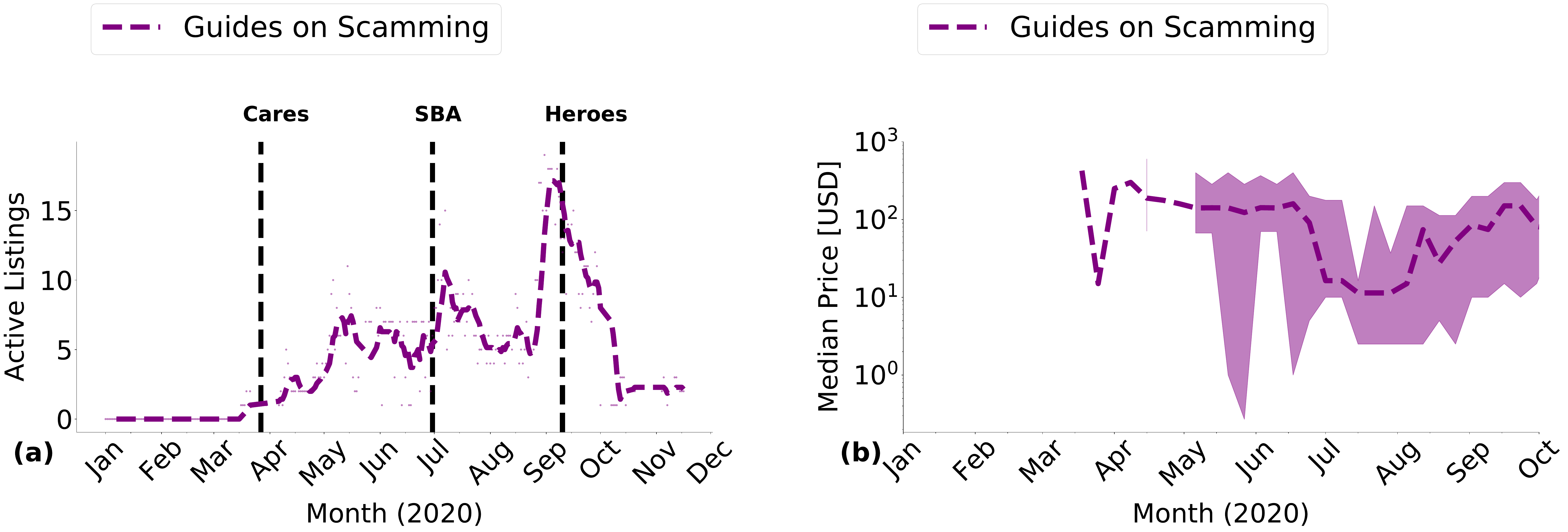}
  \caption{\edit{Time evolution of the active COVID-19 specific listings in the \textit{guides on scamming} category. (a) Seven-days rolling average of these observed listings at a given time. Black dashed vertical lines corresponded to significant COVID-19 world events, see Appendix~\ref{Timeline_covid}. (b) Seven-days median price with 95\% confidence interval for these observed listings.}}
  \label{Appendix_guides_on_scamming}
\end{figure}

\edit{In the main text, we performed a longitudinal analysis of the time evolution of all COVID-19 specific listings and all listings mentioning COVID-19, as well as the \textit{PPE} and \textit{medicines} categories, as shown in Figure~\ref{Time_series_Listings_Overall}. Now, we provide a similar analysis for COVID-19 \textit{guides on scamming}, as illustrated in Figure~\ref{Appendix_guides_on_scamming}. We observe they first appeared in March, when the first lockdown measures were adopted. The number of listings then started increasing after the Cares act was introduced in USA~\cite{cares_act}. Other peaks coincide to the extension of the SBA loan program in July~\cite{sba} and to the failing of negotiations on the Heroes act~\cite{heroes_act}, after which the number of listings decreased up to April levels. Listings in the \textit{guides on scamming} category teach people how to take advantage of several kinds of COVID-19 relief funds, or other pandemic related scam opportunities. In many western countries, new relief funds were signed on a monthly basis constant updates made on the relative listings on DWMs.}

\begin{figure}[h!]
  \centering
  \includegraphics[width=12cm]{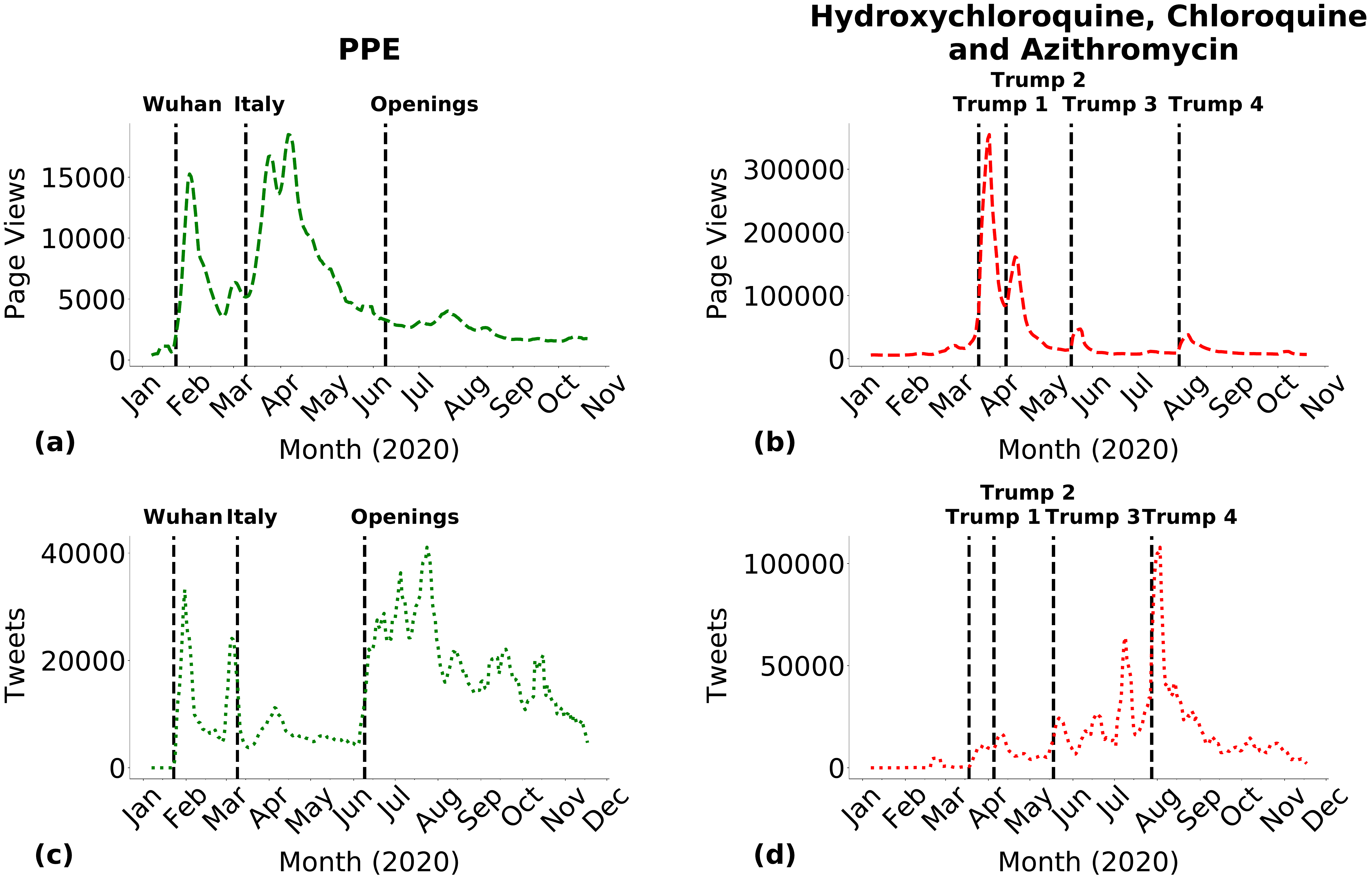}
  \caption{Wikipedia page visits for pages relative to (a) \textit{PPE}, (b) hydroxychloroquine, chloroquine and azitrhomycin. Number of tweets mentioning (c) \textit{PPE}, (d) hydroxychloroquine, chloroquine and azitrhomycin. Panels (a) and (b) corresponds to Figure~\ref{Time_series_Twitter_Wikipedia_Listings}(a) in the main text, while panels (c) and (d) to Figure~\ref{Time_series_Twitter_Wikipedia_Listings}(b). The main difference between these panels and Figure~\ref{Time_series_Twitter_Wikipedia_Listings}(a) and (b) is the linear scale on y axis.}
  \label{Appendix_time_series}
\end{figure}

In order to complement Figure~\ref{Time_series_Twitter_Wikipedia_Listings}(a) and (b) in the main text and properly show the peaks of Wikipedia page visits and tweets, we create Figure~\ref{Appendix_time_series}. The new representation of Figure~\ref{Time_series_Twitter_Wikipedia_Listings} does not modify the claims made in the main text and how major event related with COVID-19 impacted public attention. 

\begin{table}[ht]
  \centering
\caption{List of all DWMs, together with their specialization and a brief description.}
      \begin{tabular}{lll}
        \hline
          DWM & Specialization & Description \\ \hline
Atshop &	Digital Goods	& Atshop e-commerce marketplace platform\\
\multirow{2}{*}{Black Market Guns} & \multirow{2}{*}{Weapons} & Weapons Marketplace, now exit scammed \\
 &  & according to onion.live~\cite{OnionLive} \\
CanadaHQ & Mixed & Multivendor cryptocurrency marketplace \\
\multirow{2}{*}{Cannabay} & \multirow{2}{*}{Drugs} & Russian language drug marketplace \\
 &  & focusing on cannabis \\
Cannazon & Drugs (Cannabis) & Drug marketplace for cannabis products only \\
\multirow{2}{*}{Connect} & \multirow{2}{*}{Mixed} & A social network that hosts a marketplace \\
 &  & for the sale of illicit goods \\
 Cypher &	Mixed 	& Cypher is a multivendor market for the sale of drugs and digital goods. \\
\multirow{2}{*}{DarkBay/DBay} & \multirow{2}{*}{Mixed} & Multivendor cryptocurrency DWM selling \\
 &  & digital goods, drugs, and services \\
\multirow{2}{*}{Dark Market} & \multirow{2}{*}{Mixed} & Multivendor cryptocurrency DWM selling  \\
 &  & digital goods, drugs, and services \\
Darkseid & Weapons & Weapons DWM \\
\multirow{2}{*}{ElHerbolario} & \multirow{2}{*}{Drugs} & Single-vendor shop, selling just 3 products,\\
 & & primarily leaning towards Cannabis \\
\multirow{2}{*}{Empire} & \multirow{2}{*}{Mixed} & Alphabay-style DWM with BTC, LTC, XMR, \\
 &  & MultiSig, and PGP 2FA \\
 Exchange &	Mixed & Chinese language marketplace \\
\multirow{2}{*}{Genesis} & \multirow{2}{*}{Digital goods} & Marketplace selling digital identities for account \\
 &  & takeover activities\\
Hydra & Drugs & Russian language drug DWM \\
MagBO &	Digital Goods	& Shell, account and card shop\\
MEGA Darknet & Mixed & Russian language DWM \\
Monopoly &	Drugs	& Multivendor market that is primarily focused on drugs \\
\multirow{2}{*}{Mouse In Box} & \multirow{2}{*}{Digital Goods} & Marketplace that sells packages of login and session \\
 & & information acquired from web browsers with a stealer malware.\\
Plati.Market & Digital goods & digital goods DWM \\
Rocketr & Digital goods & Marketplace for the sale of illicit digital goods \\
Selly & Digital goods & Marketplace for the sale of illicit digital goods \\
Shoppy.gg & Digital goods & Marketplace for the sale of illicit digital goods \\
Skimmer Device & Skimmer devices & Marketplace selling skimmer devices \\
\multirow{2}{*}{Tor Market} & \multirow{2}{*}{Drugs} & Drug DWM focused on supplying \\
 &  & the drug marketplace in New Zealand \\
 Torrez & 	Mixed 	& Torrez is is a multivendor market that uses wallet-less payments. \\
\multirow{2}{*}{Venus Anonymous} & \multirow{2}{*}{Mixed} & Multivendor DWM selling \\
 &  & digital goods and drugs \\
White House & Mixed & Multivendor cryptocurrency DWM \\
\multirow{2}{*}{Wilhaben} & \multirow{2}{*}{Mixed} & German language DWM for the sale  \\
 &  & of illicit goods \\
Yellow Brick & Mixed & Multivendor cryptocurrency DWM \\
      \end{tabular}
\label{market_details}
\end{table}

\begin{table}[ht]
  \centering
\caption{COVID-19 related medicines appearing in the listings, together with a brief description and the number of listings related to that drug.}
      \begin{tabular}{llc}
        \hline
          Medicine & Description & Listings \\ \hline
          Hydroxychloroquine & Malaria medication & \edit{65}\\
Azitrhomycin & Antibiotic often paired with hydroxychloroquine & \edit{51} \\
\edit{Amoxicillin} & \edit{Antibiotic medication} & \edit{45} \\
Chloroquine  & Malaria medication & \edit{38} \\
\edit{Ciprofloxacin} & \edit{Antibiotic medication} & \edit{6} \\
Favipiravir & Antiviral medication used to treat influenza & 5 \\
\edit{Dxycycline} & \edit{Antibiotic medication} & \edit{4} \\
\edit{Metronidazole} & \edit{Antibiotic medication} & \edit{4} \\
Remdesivir & Antiviral medication & \edit{2} \\
Lopiravir & Antiviral medication used to treat HIV &1 \\

      \end{tabular}
\label{medicines_table}
\end{table}

\begin{table}[ht]
\centering
\caption{COVID-19 related \textit{guides on scamming} appearing in the listings, together with a brief description and the number of listings related to that sub-category.}
      \begin{tabular}{llc}
        \hline
          Topic & Description & Listings \\ \hline
          SBA loan & how to illicitly get money from the USA Small Business Loan program~\cite{sba} & 19\\
Bank account & how to exploit pandemic related security to open bank accounts & 16 \\
Fraud Pack & pack containing multiple generic covid related frauds & 7 \\
Covid-19  & Generic guides explaining how to exploit the pandemic in many different ways & 7 \\
Amazon & Amazon related fraud guides & 6 \\
GoFundMe & GoFundMe related fraud guides & 4 \\
Apple & Apple related fraud guides & 3 \\
Unemployment fund & How to illicilty get money from government unemployment funds & 3 \\
Other & Other COVID-19 related fraud guides & 34 \\

      \end{tabular}
\label{guides_table}
\end{table}

%\clearpage

\end{document}